\documentclass[epj]{svjour}
\usepackage{graphics}
\usepackage{graphicx}
\usepackage{amssymb}
\begin{document}
\title{Hybrid Compton-PET Imaging for ion-range verification}
\subtitle{A Preclinical Study for Proton-, Helium-, and Carbon-Therapy at HIT} 

\author{Javier Balibrea-Correa\inst{1} \and Jorge Lerendegui-Marco \inst{1} \and Ion Ladarescu \inst{1} \and Sergio Morell-Ortega \and Carlos Guerrero \inst{2,3} \and Teresa Rodríguez-González \inst{2,3} \and Maria del Carmen Jiménez-Ramos \inst{3,4} \and Jose Manuel Quesada \inst{2} \and Julia Bauer \inst{5,6} \and  Stephan Brons \inst{5,6} \and César Domingo-Pardo \inst{1}}

%}                     % Do not remove
%

\institute{F{\'i}sica Nuclear Experimental, CSIC-University of Valencia, Valencia, Spain, Catedrátic José Beltrán Martinez, 2, Valencia, 46980, Spain \and Departamento de Física Nuclear y Atómica y Molecular, University of Seville, Sevilla, 41012, Andalucia, Spain \and Aceleradores,Centro Nacional de Aceleradores (U. Sevilla, CSIC, Junta de Andaluc\'ia),C. Tomás Alva Edison, 7, Sevilla, 41092, Andalucia, Spain \and Department of Applied Physics II, ETSA, University of Seville, Sevilla, 41092, Andalucia, Spain \and Department Of Radiation Oncology, University Hospital Heidelberg, Heidelberg, Germany \and Heidelberg Institute For Radiation Oncology (hiro), National Center for Radiation Research in Oncology (NCRO), Heidelberg, Germany}


\date{Received: date / Revised version: date}
\abstract{
Enhanced-accuracy ion-range verification in real time shall enable a significant step forward in the use of therapeutic ion beams. Positron-emission tomography (PET) and prompt-gamma imaging (PGI) are two of the most promising and researched methodologies, both of them with their own advantages and challenges. Thus far, both of them have been explored for ion-range verification in an independent way. However, the simultaneous combination of PET and PGI within the same imaging framework may open-up the possibility to exploit more efficiently all radiative emissions excited in the tissue by the ion beam. Here we report on the first pre-clinical implementation of an hybrid PET-PGI imaging system, hereby exploring its performance over several ion-beam species (H, He and C), energies (55 MeV to 275 MeV) and intensities (10$^7$-10$^9$ ions/spot), which are representative of clinical conditions. The measurements were carried out using the pencil-beam scanning technique at the synchrotron accelerator of the Heavy Ion Therapy centre in Heidelberg utilizing an array of four Compton cameras in a twofold front-to-front configuration. The results demonstrate that the hybrid PET-PGI technique can be well suited for relatively low energies (55-155 MeV) and beams of protons. On the other hand, for heavier beams of helium and carbon ions at higher energies (155-275 MeV), range monitoring becomes more challenging owing to large backgrounds from additional nuclear processes. The experimental results are well understood on the basis of realistic Monte Carlo (MC) calculations, which show a satisfactory agreement with the measured data. This work can guide further upgrades of the hybrid PET-PGI system towards a clinical implementation of this innovative technique.}

\PACS{
      {hadron therapy}{} \and {cancer}{} \and {ion range monitoring}{} \and {PGI}{} \and {PET}{} \and {hybrid system}{} % end of PACS codes
} %end of abstract

\maketitle

\section{Introduction}\label{sec1}
Hadron-therapy represents a well established technique for cancer treatment with good prospects for further improvements and extension to other diseases in the future~\cite{Durante:2016}. State-of-the-art treatments aim at maximum dose deposition in the tumor area thanks to the large energy loss of the ions at the end of their track along the Bragg peak~\cite{Wilson:46,Knopf:13}. This is particularly relevant to minimize damage in healthy tissues surrounding the tumor, thus reducing as well long term secondary effects. This methodology is therefore especially well-suited for many pediatric cases and tumors close to sensible organs~\cite{Knopf:13}.

However, a full benefit of therapeutic hadron beams is still hold back due to uncertainties in the ion range or penetration depth in tissue. Present treatment plans are based on conversion of photo-attenuation coefficients into ion stopping power. Thus, semi-empirical relative stopping powers in tissue are approximated from X-ray computed tomography measurements (either in water or in tissue), which leads to range uncertainties of 1-3\%. In turn, such uncertainties impose conservative safety margins of up to 3.5\%+3~mm~\cite{Moyers:2010,Paganetti:2012,Yang:2012}, meaning that normally a significant region of healthy tissue is irradiated in order to ensure that the full tumor volume is treated. This situation is particularly critical for the treatment of tumors in the neighbourhood of sensible organs.

New methods for accurate ion-range verification could help to improve this situation~\cite{Knopf:13}. In particular, real-time sub-mm range verification~\cite{Paganetti:2012,Kraan:15} would allow one to enhance the spatial accuracy in dose delivery distribution, thus increasing correspondingly the benefits of this technique and even extend it to other diseases such as ventricular tachycardia and other cardiovascular disorders~\cite{Durante:2016,Durante:2017,Durante:2019}. Presently, there are two gamma-ray imaging techniques that have been thoroughly investigated and researched in clinical conditions for ion-range verification, Positron-Emission Tomography (PET) and Prompt-Gamma Imaging (PGI). 

PET was the first methodology for online monitoring~\cite{Llacer:78} and it is one of the most extensively researched and applied techniques for clinical ion-range verification~\cite{Enghardt:04,Bisogni:2016}. Ion-range verification via PET is commonly based on $\beta^+$ emitters (predominantly $^{11}$C, $^{13}$N and $^{15}$O~\cite{Kraan:14,Moteabbed:2011,PARODI:2023}) produced as a consequence of nuclear reactions induced by a primary stable ion-beam along the irradiated tissue. There are four PET protocols for ion-range monitoring: in-room, off-line, in-beam, and inter-spill modes~\cite{Kraan:15}. The main challenges for the former two are related to biological wash-out effects~\cite{Knopf:13,PARODI:2007}, while the latter two methodologies have to deal with relatively large backgrounds and reduced signal sensitivity\cite{Shakirin:2011}. Furthermore,  these approaches require the use of innovative non-standard PET systems to mitigate and account for such effects~\cite{Bisogni:2016,Ferrero:2018}. Still, exploiting short-lived $\beta^+$ emitters~\cite{Dendooven:2015}, such as $^{12}$N (T$_{1/2}=10$~ms), it has been demonstrated that in-beam PET imaging may provide real-time sensitivity to ion range-shift variations of better than 2~mm~\cite{Ozoemelam:2020a,Ozoemelam:2020b}. In general, detection statistics for real-time monitoring using 511 keV $\gamma$-rays becomes quite challenging owing to the low production yields for the short-lived $\beta^+$ emitters, the limited signal-to-background ratios and the increasing contribution from long-lived positron emitters produced at other scanning layers along the treatment~\cite{Bisogni:2016}. 
Recently, also the idea of using  primary $\beta^+$-unstable ion-beams~\cite{Llacer:1988}, like $^{15}$O, for simultaneous treatment and range monitoring has been experimentally demonstrated~\cite{Purushothaman:23}. Finally, a common aspect of all PET-based methodologies is the fact that the high intrinsic spatial resolution of PET imaging is counterbalanced by the relatively broad spatial distributions of the $\beta^+$ emitters, especially with proton- and light-ion beams\cite{Kraan:15,PARODI:2023}.

PGI was first proposed by Stichelbaut and Jongen~\cite{Stichelbaut:03}, and it was soon demonstrated by Min et al.~\cite{Min:2006} utilizing a mechanically collimated gamma-camera. PGI thus relies on radiative nuclear reactions occurring as the ions slow down along the patient tissues~\cite{Bom:2012}. This quasi- instantaneously emitted secondary radiation consists mainly of $\gamma$-rays in a range covering up to 5-6 MeV, and beyond~\cite{Krimmer:18}. Thus, from a conceptual point of view, these $\gamma$-rays are especially well suited for real-time monitoring due to the high spatial and temporal correlation with the primary proton range. However, from an experimental standpoint, PGI becomes also very challenging due to the requirement of in-beam measuring conditions, which include very large gamma-ray and neutron-induced backgrounds~\cite{Min:06,Testa:08,Verburg:2014} and very high instantaneous count-rate requirements. PGI real-time monitoring is also constrained by the low efficiency radiation detectors at high $\gamma$-ray energies~\cite{Smeets:2012,Perali:2014,Polf:2015,Munoz:2021} and overall detector performance at such high count rates~\cite{Knopf:13}.
%and ion interactions with the accelerator gantry~\cite{Ortega:2015}.
%Different approaches are currently under development for real-time ion range verification relying on prompt $\gamma$-rays, mainly based on PG timing~\cite{Golnik:2014,Hueso:2015}, $\gamma$-ray spectroscopy~\cite{Verburg:2014,Hueso:2018} and PGI~\cite{Peterson:2010,BARRIENTOS:2024}. 
A slit-camera has been developed and extensively used in clinical treatments with very satisfactory results~\cite{Smeets:2012}. However, apart from its spatial 1D-sensitivity, slit-cameras have a rather low detection efficiency ($\sim$10$^{-5}$), which severely constrains their precision for the determination of the Bragg-peak falloff in the 1D-profile~\cite{Krimmer:18}. This drawback can be overcome by means of electronic collimation or Compton imaging techniques, which further enable a 2D- (or even 3D-) spatial sensitivity~\cite{Schonfelder:1973}.  Despite promising recent results~\cite{Taya:2016,Draeger:2018,BARRIENTOS:2024} there are still remarkable limitations before these systems can be routinely used in clinical treatments. Such limitations are mainly related to the broad gamma-ray energy distributions, the large gamma-ray and neutron-induced backgrounds and the very high instantaneous count rates~\cite{Knopf:13,Krimmer:18}.
%%%%%%%%%%%%%%%%%%%%%%%%%%%%%%%%%%%%%%%%%%%%%%%%%%%%%%%%%%%%%%%%%%%%%%%%%%%%%%%%%%%%%%%%%%%%%%%%%%%%%

Both PGI and PET imaging could be simultaneously combined in the so-called hybrid scheme, as proposed by K. Parodi in 2016~\cite{Parodi:16}. Such an hybrid imaging system would enable the possibility to combine the functional and tomographic inherent features of PET and its high intrinsic spatial resolution, with the high-yield prompt $\gamma$-rays that are closely linked in time and position to the Bragg peak. As suggested by Lang~\cite{Lang:14}, this idea could be implemented by adapting systems based on multiple Compton cameras intended for high-sensitivity three-$\gamma$-ray correlations. Alternatively, the Krakow group has recently explored the possibility to extend their multi-photon PET scanner (J-PET)~\cite{Moskal:2021a,Moskal:2021b} also for hybrid PET-PGI with promising expectatives~\cite{Kozani:2023}. 
%
%The hybrid PET-PGI technique is expected to be well suited for pulsed beam structures, where PGI can be exploited during spill delivery, whereas PET-imaging could be accomplished in the short time-intervals between spills~\cite{Balibrea22a,Balibrea22b}. 

In a previous work~\cite{Balibrea22a,Balibrea22b} we conducted a proof-of-concept experiment to demonstrate the feasibility of hybrid PET-PGI utilizing the 18~MeV Cyclotron radiobiological research line at the Centro Nacional de Aceleradores (CNA) in Seville~\cite{BARATTOROLDAN:2020}. In that work at CNA, following the multi-Compton arm approach~\cite{Lang:14}, the experimental setup consisted of two Compton cameras in front-to-front configuration, thus enabling both Compton PGI and PET imaging at the same time. These Compton imagers were initially designed for nuclear astrophysics experiments~\cite{DOMINGOPARDO:2016} and, once fully developed, utilized for neutron-capture time-of-flight experiments at CERN n\_TOF~\cite{Lerendegui:2023}. Thus, their unconventional geometry design with one large scatter- and four large absorber detectors (1S+4A)~\cite{Babiano:20} was optimized to maximize detection efficiency for $\gamma$-rays of energies up to 5-6~MeV over a large field-of-view (FOV)~\cite{Babiano:20}. Other requirements included high count-rate capabilities and low sensitivity to neutron induced-background~\cite{Babiano:20,Domingo:2023}. The experiment carried out at CNA with the hybrid system confirmed a sub-mm position accuracy and a fully consistent PET and PGI position reconstruction. However, those measurements were performed at only 18 MeV proton energy, and the beam-time structure was quite different from the one available in clinical treatments.

In this article we present new measurements carried out with the hybrid imaging system under preclinical ion-beam conditions at the Heidelberg Ion Therapy (HIT) center~\cite{Combs:10}, utilizing beams of protons, He- and C- ions at clinical intensities (10$^{7}$-10$^{9}$ ions/spot) and with a broad range of energies (55-225 MeV). The hybrid imaging system was similar to the one utilized before at CNA, but upgraded with two additional Compton cameras (four in total) along the beam axis, thereby covering a large FOV for simultaneous in-situ PGI and PET imaging. The primary objectives of the experiment were to demonstrate the feasibility of our hybrid in-beam PGI-PET setup for ion-range verification under preclinical beam conditions, to determine the experimental sensitivity of the hybrid system and to gather technical information for future potential upgrades.

\section{Methods}\label{sec2}
The measurements reported here were carried out at the experimental line of the Heidelberg Ion-Beam Therapy Center (HIT)~\cite{Combs:10}. The hybrid PGI-PET setup consisted of four Compton imagers, also called i-TED modules~\cite{DOMINGOPARDO:2016,Babiano:20}. They are referred in this work as i-TED -A, -B, -C and -D. Each individual imager is made of the largest commercially available LaCl$_{3}$(Ce) monolithic scintillation crystals optimized to cover a wide range of $\gamma$-ray energies, from few hundreds of keV up to several MeV~\cite{DOMINGOPARDO:2016,Babiano:20,Domingo:2023}. The scatter (S) detector in each imager consists of one 50$\times$50$\times$15~mm$^{3}$ crystal, whereas four co-planar crystals (4$\times$A) with a size of 50$\times$50$\times$25~mm$^{3}$ each were utilized for the absorber plane of each imager~\cite{Babiano:20}. Only i-TED C  had a scatter crystal that was 5~mm thinner than the others (50$\times$50$\times$10~mm$^{3}$). Each monolithic LaCl$_{3}$(Ce) crystal was optically coupled to a 8$\times$8 pixels silicon photomultipliers (SensL ArrayJ-60035-65P-P). The detector signals are readout and processed using an acquisition system based on PETsys Front-End Board D version 2 (FEB/D-1024)~\cite{DIFRANCESCO:2016}. The latter was also used for applying a voltage bias to the SiPMs.

\begin{figure*}
    \centering
    \includegraphics[width=0.7\textwidth]{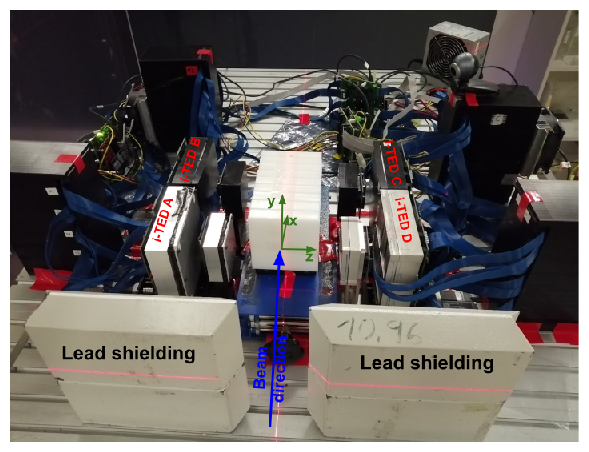}
    \caption{Photograph of the hybrid PET-PGI system made of four Compton imagers in two-fold front-to-front configuration during the position sensitivity study for different incident incoming particles.}
    \label{fig:Exp_setup}
\end{figure*}

The four Compton cameras were placed in the experimental area on opposite sides of the ion-beam axis, as shown in Fig.~\ref{fig:Exp_setup}. The reference axis of the experimental setup are drawn in the same figure.
%: the negative $x$-axis corresponds to the direction of the incoming ion beam, the $y$-axis is aligned with the height of the i-TED imagers, and the $z$-axis, perpendicular to the other two, runs from i-TED A to i-TED D. 
The origin of the coordinate system ($x$=0 mm, $y$=0 mm) corresponds to the geometric center of i-TED cameras A and D, while $z$=0 mm corresponds to the position of the proton beam incident axis. This reference system, along with the origin of coordinates, provides an absolute coordinate framework for Compton and PET image reconstruction. The distance between the front faces of the i-TED scatter cameras, i.e., i-TED -A and -D, and i-TED -B and -D, was 198(1) mm. The lateral space between consecutive i-TED cameras, i.e., i-TED -A and -B, -C and -D, was 10 mm. Lead blocks with a thickness of 50 mm were placed between the exit of the accelerator and the i-TED Compton cameras to shield them from any potential background contributions related to ion interactions with the accelerator gantry. The  experimental setup implemented here was a natural evolution of the one used in our previous work at CNA-Seville, where we investigated hybrid PET-PGI using two single i-TED imagers in front-to-front configuration with a proton-beam of 18 MeV~\cite{Balibrea22a,Balibrea22b}. It is worth noting that the set-up implemented at HIT was still suboptimal in terms of efficiency. A factor of $\sim$2 higher efficiency could have been achieved with a cross-geometry configuration, such as the one shown in Ref.~\cite{Lerendegui:2022}. However, the extended FOV of the configuration implemented at HIT allowed us to get the full $\gamma$-ray picture along the entire phantom volume. As discussed later, this approach was of relevance for understanding and interpreting the results in a consistent manner.

Default treatment conditions were used throughout the entire experimental runs, hereby emulating a realistic clinical environment with the aim of demonstrating the hybrid PGI-PET capabilities of our experimental setup. The beam intensity was of 10$^{7}$ protons per spill, 4~mm spatial width at 155 MeV (smaller at higher beam energy), and synchrotron periods of 300 ms, divided into a 45 ms beam spill and a 255 ms break. The accelerator provided a trigger signal each time a spill was delivered to the experimental area, thereby establishing a beam-related time framework for the acquisition system and offline analysis. 

Two different types of phantoms or targets were utilized along this work. A cylindrical graphite target with 25 mm diameter and 50 mm length was placed at different controlled positions along the entire PET field of view. A summary of the irradiation characteristics used for this part of the experiment is provided in Tab.~\ref{tab:Spills_And_Configuration}. 
In the second part of the experiment, 50$\times$50$\times$180~mm$^{3}$ polyethylene blocks were mounted on a high-precision linear positioning stage (M683 from PI-miCos). This device has a load capacity of 50 N and includes an integrated linear encoder with a resolution of 0.1 $\mu$m. The embedded piezoceramic linear motor allowed us to control the position of the phantom with sub-micrometric precision. The target and linear device were placed within the PET field of view, aligned with the beam direction. The experiment was repeated with protons, He-ions, and C-ions at 155, 155 and 275 MeV, respectively. The details of the different irradiations are provided in Tab.~\ref{tab:Spills_And_Configuration}. After each individual irradiation, the polyethylene (PE) block was replaced to remove remnant $\beta^{+}$ isotopes from the previous irradiation, which could otherwise interfere with the in-spill and off-spill PET results and conclusions.

\begin{table*}[htb!]
    \centering
    \begin{tabular}{c c c c c c }\hline
     Beam & Energy [MeV] & phantom & Displacement [mm]& duration [s] & Number of Spills \\ \hline \hline
    p & 55 & Graphite & 0(1) & 840 & 475\\ 
     & & Graphite & -60(1) & 780 & 467\\
     & & Graphite & -120(1) & 840 & 470\\ \hline
   p &  155 &  PE & 0.0(1) & 840 & 164\\
     & &  PE & 1.00(1) & 2040 &  573\\
     & &  PE & 1.50(1) & 1800  & 1270\\ \hline 
   He-ions &  155 &  PE & 0.00(1) & 2040 & 1068\\
     & &  PE & 1.00(1) & 1980 &1103\\ \hline
   C-ions &  275 &  PE & 0.00(1) & 1920 & 1004 \\
     & &  PE & 1.00(1) & 2040 &  1003 \\ \hline\hline
    \end{tabular}
    \caption{Summary of the irradiations and characteristics performed during the experimental campaign.}
    \label{tab:Spills_And_Configuration}
\end{table*}

The details of the different measurements described in Tab.~\ref{tab:Spills_And_Configuration} include the phantom type used, displacement from the reference position, time duration, and number of spills used to reconstruct both PGI- and PET- images. It is worth to mention that due to a temporal synchronization issue between the acquisition system and accelerator trigger signal, there is a large difference between the number of spills used for the sensitivity study with protons. The issue was solved afterwards for He- and C- ions as it is reflected in the same table.

\begin{figure}
    \centering
    \includegraphics[width=0.5\textwidth]{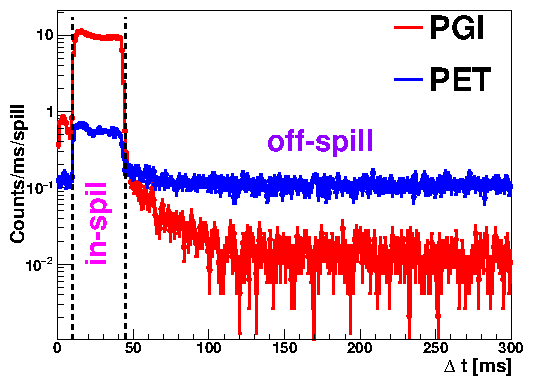}
    \caption{Average registered coincidence count-rate for PGI- and PET- imaging during the delivery of the beam (in-spill) and waiting time (off-spill) during the the 55 MeV proton beam irradiation on the graphite target as a function of spill time. A similar plot can be extracted from the second part of the measurement}
    \label{fig:DutyCycle}
\end{figure}

An example of the average PGI and PET coincidence count rates as a function of time is shown in Fig.~\ref{fig:DutyCycle}, represented by the red and blue lines, respectively. One of the most challenging aspects for PGI is the high count rates during spill. Instantaneous Compton PGI in-spill counting rates are of 10 kHz (in time-coincidence between S- and A-detectors), which is a factor of 10 (100) larger compared with PET in-spill (off-spill). It is worth noting that absolute count rates per detector are about one order of magnitude higher in both cases. 

The methodology and algorithms for data analysis have been described in detail in previous works~\cite{BALIBREACORREA:2021,BALIBREA2021b,Balibrea22a,Balibrea22b}. A brief summary is given here for completeness. The 3D $\gamma$-ray hit position reconstruction in each position-sensitive detector (PSD) was implemented following the methodology described in Ref.~\cite{BALIBREACORREA:2021}. The individual PSD detectors were  calibrated in energy using a standard $^{152}$Eu calibration source, as well as the 511~keV and 4.4~MeV $\gamma$-rays, the first- and the second-escape peaks measured during the proton irradiation on the graphite target. This approach ensured a complete energy calibration over the full energy range of interest. Compton imaging with the individual i-TED cameras was accomplished through $\gamma$-ray hit events detected in time coincidence ($\Delta t^{PGI}_c = 10$ ns) between the scatter-PSD and any of the four rear PSDs. Deposited-energy selections were applied to the sum of the event energies, ranging from 600 keV to 6 MeV. At least 600 keV of deposited energy in the absorber plane was required to accept the event. In this way, $e^{+}$ decays from pair production and random coincidences with other background sources were significantly reduced. The Compton images were obtained by means of our own implementation of the analytical inversion algorithm based on a spherical polynomial expansion published by Tomitani and Hisarawa in 2002~\cite{Tomitani:02}. To speed up the calculation, the algorithm was implemented on a GPU using the CUDA 11.1 toolkit~\cite{Lerendegui:2022}. The image plane was positioned at the proton beam axis, i.e., $z$ = 0 mm. Individual Compton images from the different i-TED imagers were then combined into a single image, thereby taking into consideration the efficiency and Compton field of view (FoV) of each i-TED imager. The latter parameters were calculated using MC simulations, as  explained below. PET imaging was accomplished through time-coincidence events ($\Delta t^{PET}_c = ??$ ns) between any PSDs on opposite sides of the proton beam axis. This includes coincidences between scatter-scatter, scatter-absorber, and absorber-absorber detectors from different i-TED imagers. Only events with detected energies (the sum of the registered energy in the individual detectors) in the range of 0.9 to 1.1 MeV were accepted for imaging. In this case, PET images were reconstructed using a simple analytical algorithm, where straight lines of response (LOR) between the $\gamma$-ray interaction 3D positions at each PSD detector were intersected with the central axial plane, aligning with the Compton image plane at the proton beam axis. As with the Compton technique, the PET images were reconstructed at $z$ = 0 mm and corrected for efficiency based on calculations from MC simulations.

\begin{figure}
    \centering
    \includegraphics[width=0.5\textwidth]{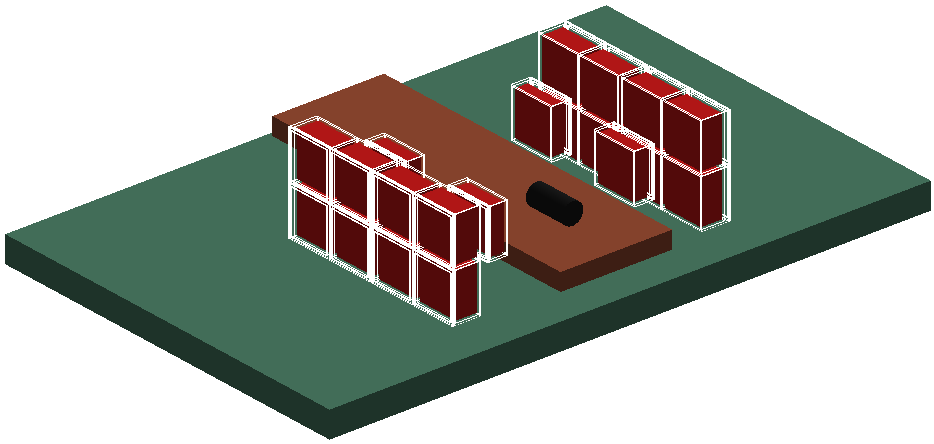}
    \caption{Geometry of the experimental setup as implemented in the \textsc{Geant4} MC simulations. See text for details.}
    \label{fig:MC-setup}
\end{figure}

A simplified geometry of the experimental setup was implemented in a C++ MC application based on the \textsc{Geant4} toolkit, version 4.10.7.p03~\cite{ALLISON:2016}. The MC simulation included the standard electromagnetic package option 4, as well as the radioactive decay and hadron-therapy simulation packages commonly used~\cite{Verburg:2012}. Various scenarios were simulated to match the experimental configurations: one with the graphite target in place and a second with a polyethylene phantom. Fig.~\ref{fig:MC-setup} shows the MC setup, including the graphite target. Energy and position resolutions were applied to the detected events using parameters adjusted from an experimental characterization~\cite{Balibrea22b}. A picture of the implemented geometry is shown in Fig.~\ref{fig:MC-setup}, depicting the scenario with the graphite target in place. The Compton field of view (FoV) and efficiency for the individual i-TED imagers were calculated by simulating individual $\gamma$-rays ranging from 100 keV to 10 MeV and applying the same analysis conditions as used for the experimental data. The large polyethylene phantom was used in the MC setup for this calculation. From the simulations, i-TED C showed 7\% lower efficiency for Compton imaging compared to the other i-TED imagers, which was expected from the 5~mm thinner scatter crystal in this imager. This effect was corrected in the experimental images to ensure consistent results. The Compton FoV helps to reduce artifacts produced by the ambiguity in Compton imaging reconstruction.

\section{Results}

The experiment was divided in two parts, each one corresponding to a different type of phantom and goals. In the first part a proton beam of 55~MeV impinged a cylindrical graphite target, which was placed at three different positions along the beam axis. 
Pictures of the experimental setup taken during this set of measurements are shown in the three top panels of Fig.~\ref{fig:fotos}. The second set of measurements involved a large and movable polyethylene phantom, as shown in Fig.~\ref{fig:setup_PE}, and higher beam energy up to 155~MeV for protons and He-ions, and 275~MeV for C-ions. The PE-target was placed with a precision of a few $\mu$m, as described in Sec.\ref{sec2}. 

Beam intensities similar to regular clinical treatments at HIT were maintained throughout the entire experimental campaign with values of about  10$^{7}$protons/spill. For an optimal exploration of the hybrid PET-PGI technique a 15\% duty-cycle was utilized, with pulsed-beam periods of 300~ms and 45~ms beam-spill delivery. The unconventional pulsed-beam structure and the time-interval or break between spills of 255 ms (see Fig.~\ref{fig:DutyCycle} in  Sec.\ref{sec2}) enabled the study of range verification via both PET in-spill and PET off-spill. PGI 2D-diagrams were reconstructed from the data collected in-spill, while PET images were reconstructed for both, in-spill and off-spill data. 
%Beam conditions similar to regular clinical treatments at HIT were maintained throughout the entire experimental campaign, including intensities (10$^{7}$protons/spill) and synchrotron periods of 300~ms with a 15\% duty-cycle consisting of a 45~ms beam-spill delivery and a time-interval or break between spills of 255 ms (see Fig.~\ref{fig:DutyCycle} in Methods section). PGI 2D-diagrams were reconstructed from the data collected in-spill, while PET images were reconstructed for both, in-spill and off-spill data. 

\subsection{Hybrid PET-PGI with 55 MeV protons}

\begin{figure*}
\centering
\begin{tabular}{c c c}\hline
0 mm & -60 mm & -120 mm \\ \hline \hline
\includegraphics[width=0.3\textwidth]{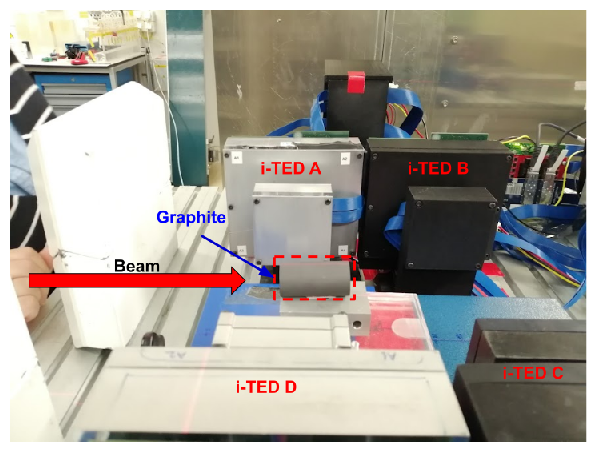} &
\includegraphics[width=0.3\textwidth]{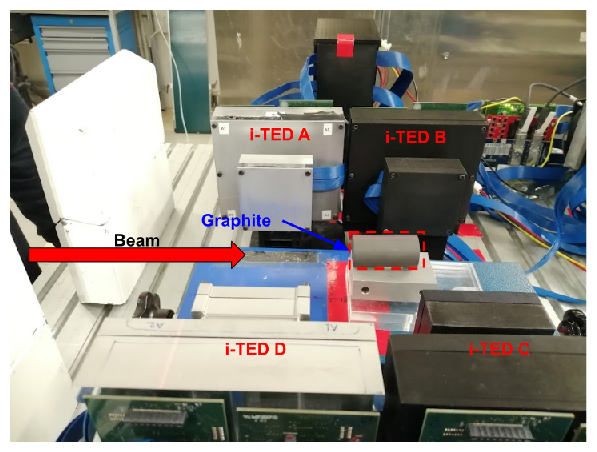} &
\includegraphics[width=0.3\textwidth]{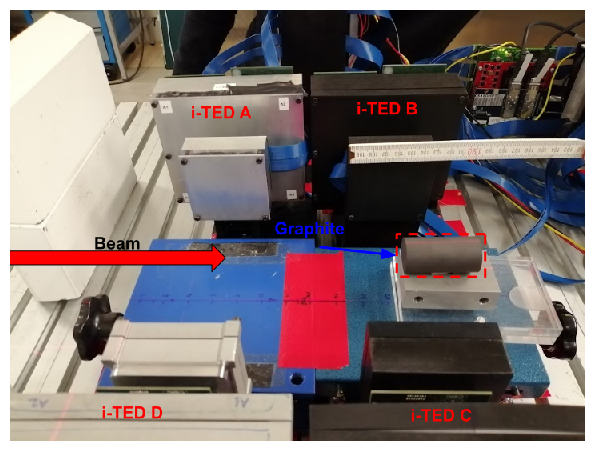} \\
\includegraphics[width=0.33\textwidth]{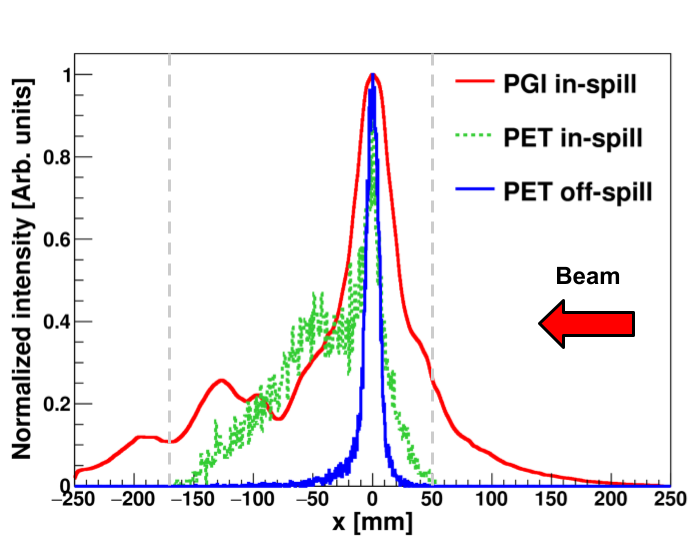} &
\includegraphics[width=0.33\textwidth]{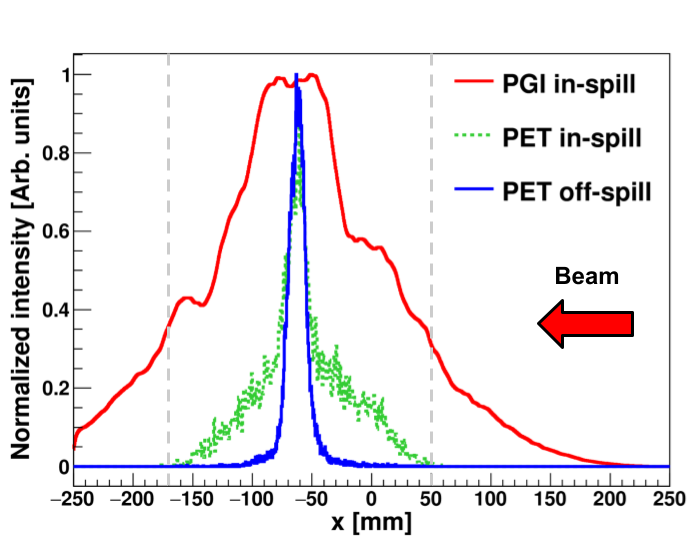} &
\includegraphics[width=0.33\textwidth]{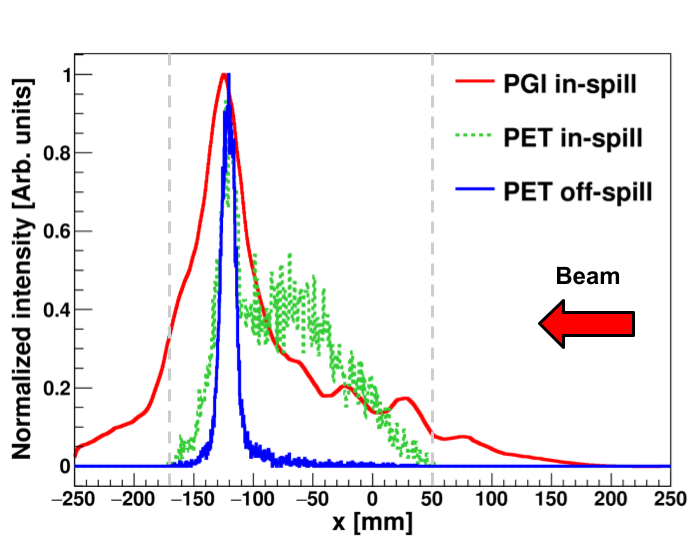} \\
\end{tabular}
\caption{Graphical summary for the hybrid PET-PGI at 55~MeV proton energy in a graphite cylinder placed at three different positions along the beam axis. Top rows: Pictures of the experimental setup with the phantom shifted in intervals of 60~mm. The red arrow indicates the beam direction and the graphite position has been highlighted with a red dashed line. Bottom raw: 1D projections of 2D-reconstructed PGI, and 1D projections of the 2D-reconstructed in-spill and off-spill PET distributions. The dashed vertical gray lines indicate the PET FoV.}% See also Fig.~\ref{fig:validation}.
\label{fig:fotos}
\end{figure*}

The graphite target consisted of a cylinder with a diameter of 25 mm and a length of 50 mm (see Fig.\ref{fig:fotos}). This material was selected for the first measurements at low proton energy due to its high stopping power and significant yields for prompt-gamma and $\beta^{+}$ production~\cite{Horst:2019,Balibrea22b,RODRIGUEZ:2023}. It ensured a substantial in-spill coincidence rate between scatter and absorber planes in each Compton imager, as well as a large number of 511 keV $\gamma$-ray coincidence events between Compton imagers for both in-spill and off-spill PET imaging. Additionally, at this beam energy, both the $\beta^+$ production region and the Bragg peak in graphite are expected to be relatively narrow and coincident. As a result, the data for PGI, in-spill, and off-spill PET can be directly compared using 2D-reconstructed images and their 1D projections along the beam axis.

The irradiation field with the pencil-beam scan consisted of a single spot, centered in the transverse plane of the target with the aid of a calibrated laser system. All irradiation points were made at the same depth (55 MeV energy), each lasting slightly more than 2 minutes and comprising, on average, 470 spills per target position (see Tab.~\ref{tab:Spills_And_Configuration} in Sec.\ref{sec2}). Thus, an average of 4$\times$10$^{9}$ protons per irradiation was available for Compton PGI and PET imaging reconstruction. The irradiations were conducted at three different positions, spaced 60(1) mm apart along the beam axis, as shown in the rows of Fig.~\ref{fig:fotos}. The bottom panels in the latter figure show the corresponding 1D PGI diagram projection (red), along with the in-spill (green) and off-spill (blue) in- and off- spill PET 1D distributions. All 1D reconstructed distributions were normalized to the maximum. It is worth noting that the orientation of the reconstructed images is inverted relative to the corresponding photographs. 

The first remarkable aspect from the reconstructed results concerns the very narrow, clean and position-sensitive PET distributions. The short time-intervals between spills naturally enhanced the contribution from short-lived $\beta^{+}$ emitters, such as $^{13}$O ($t_{1/2}\sim$8 ms) $^{12}$N ($t_{1/2}\sim$10 ms), and $^{9}$C ($t_{1/2}\sim$126 ms), with respect to long-lived isotopes such as $^{10,11}$C $^{13}$N and $^{15}$O with $t_{1/2}$ in the order of seconds or minutes. Indeed, no background-subtraction was deemed necessary for either PET and PGI images. The in-spill PET distributions clearly show a superposition of a narrow-peak and a broad contribution. The former matches well with the one obtained off-spill, as well as with the maxima of the PGIs. In fact, the underlying broad distribution remains constant in position and it can be most probably ascribed to the bulk of pair-production and $\beta^{+}$-emitters produced in-beam along the target volume~\cite{Knopf:13}. The constant position and shape of the broad distribution may indicate that at least a significant part of positron-annihilation events are taking place in the sensitive volume of the detectors themselves during the in-spill interval of 45~ms. 

The maxima of both PGI and PET (in- and off-spill) 1D projections are reported in Tab.\ref{tab:results_55MeV}. Maximum deviations of 4.2 mm between PET and PGI are found at -60~mm position, while differences of only 0.8 mm are obtained between in- and off-spill PET. 
%The differences can be attached to small miss-positioned in the alignment of the individual Compton imagers along the beam line. 
The width or resolution for PGI is between a factor of 2 and 5 larger than the one obtained with PET, owing to the worse intrinsic angular resolution for the Compton technique when compared to PET~\cite{Babiano:20,Balibrea22b}. It is worth noting that the PGI distribution is relatively broad for -60~mm position, when compared to positions at 0 and -120~mm, respectively. This effect is to be ascribed to the larger Compton angles subtended for that configuration for all four imagers (see pictures in Fig.~\ref{fig:fotos}).

\begin{table*}[h]
    \centering
    \begin{tabular}{c c c c}\hline
    Graphite Position [mm] & Method & Maximum position [mm] & $\sigma$ [mm] \\ \hline
             0          & PGI       &  0(9)   &  20(40)  \\
                       & PET in-spill   & -0.1(3) &  8.7(5)  \\
                      & PET off-spill  & -0.16(5) &  5.43(5)\\ \hline
             -60         & PGI       & -66(9)  &  50(120)  \\
                       & PET in-spill   & -62.3(2) & 9.8(4)  \\
                      & PET off-spill  & -61.8(2) & 7.1(2) \\ \hline
             -120         & PGI       & -124(10)  & 18(50)  \\
                       & PET in-spill   & -120.3(3) & 8.3(4)\\
                      & PET off-spill  & -121.1(7) & 5.8(7) \\ \hline
    \end{tabular}
    \caption{1D PGI in- and off-spill PET maximum position and resolution calculated from bottom rows of Fig.~\ref{fig:fotos}.}
    \label{tab:results_55MeV}
\end{table*}

\begin{figure*}[!htbp]
\centering
\begin{tabular}{|c|c|c|}\hline
0 [mm] & -60 [mm] & -120 [mm] \\ \hline \hline
\includegraphics[width=0.33\textwidth]{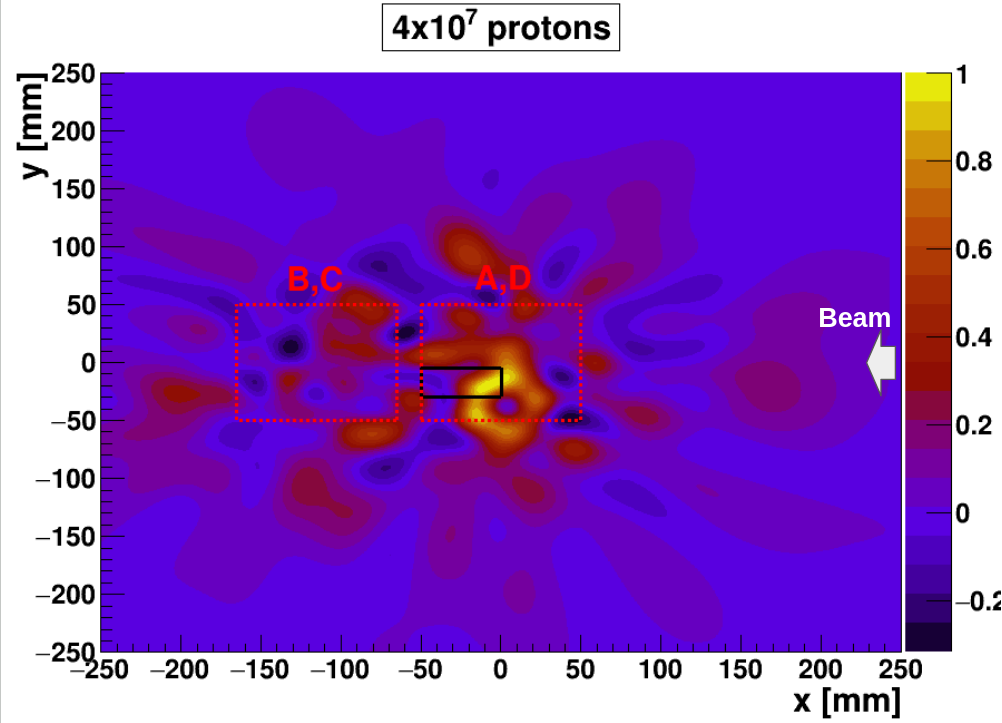} &
\includegraphics[width=0.33\textwidth]{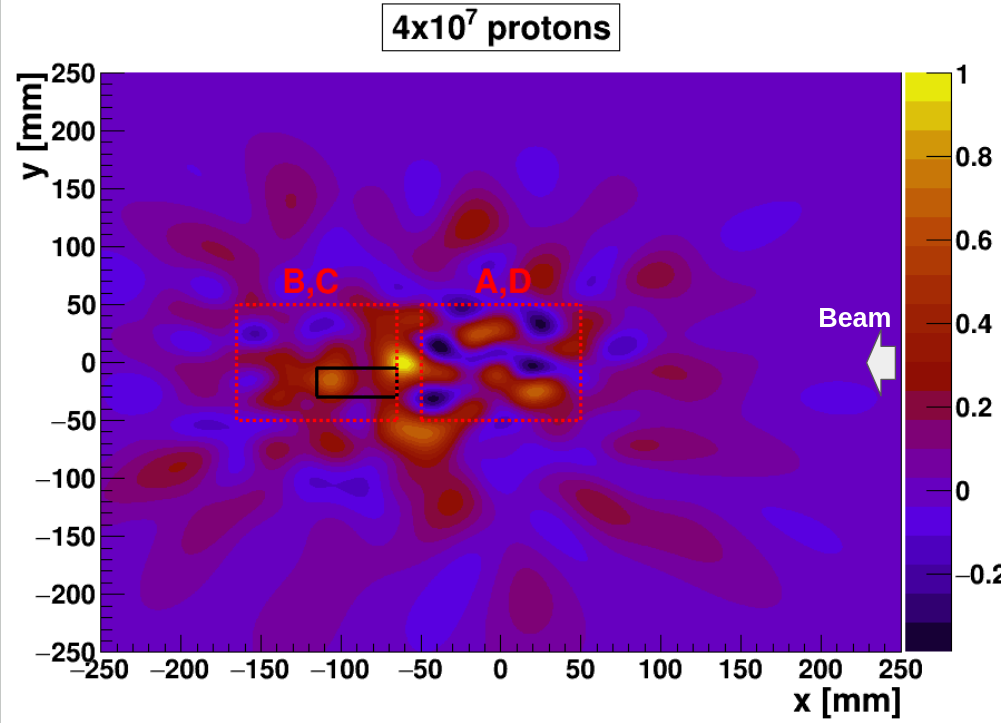} &
\includegraphics[width=0.33\textwidth]{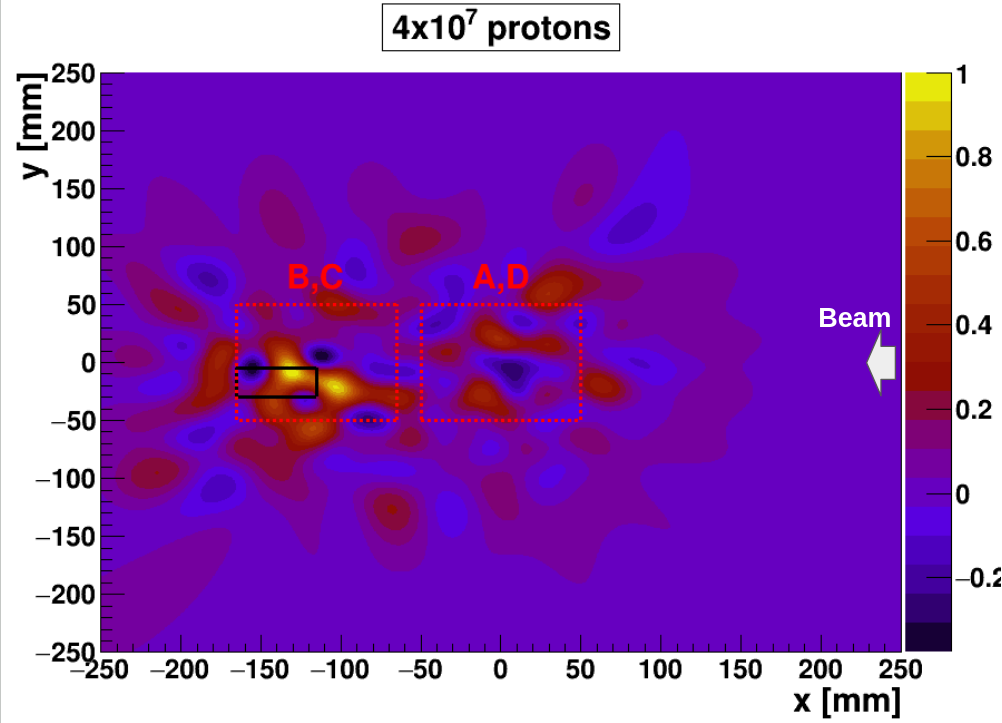} \\
\includegraphics[width=0.33\textwidth]{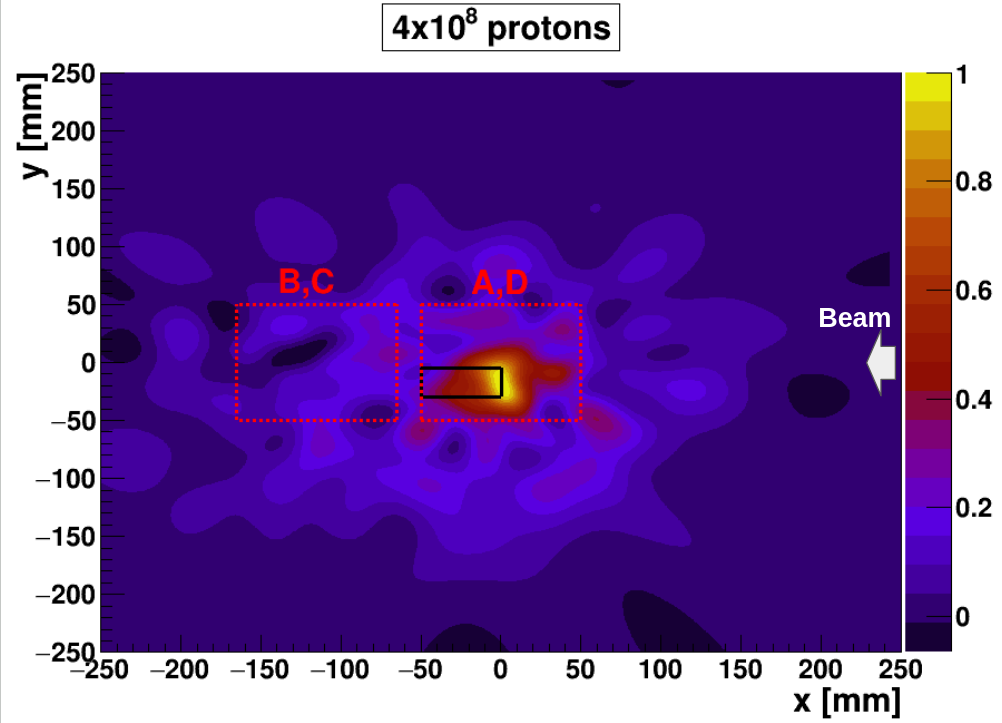} &
\includegraphics[width=0.33\textwidth]{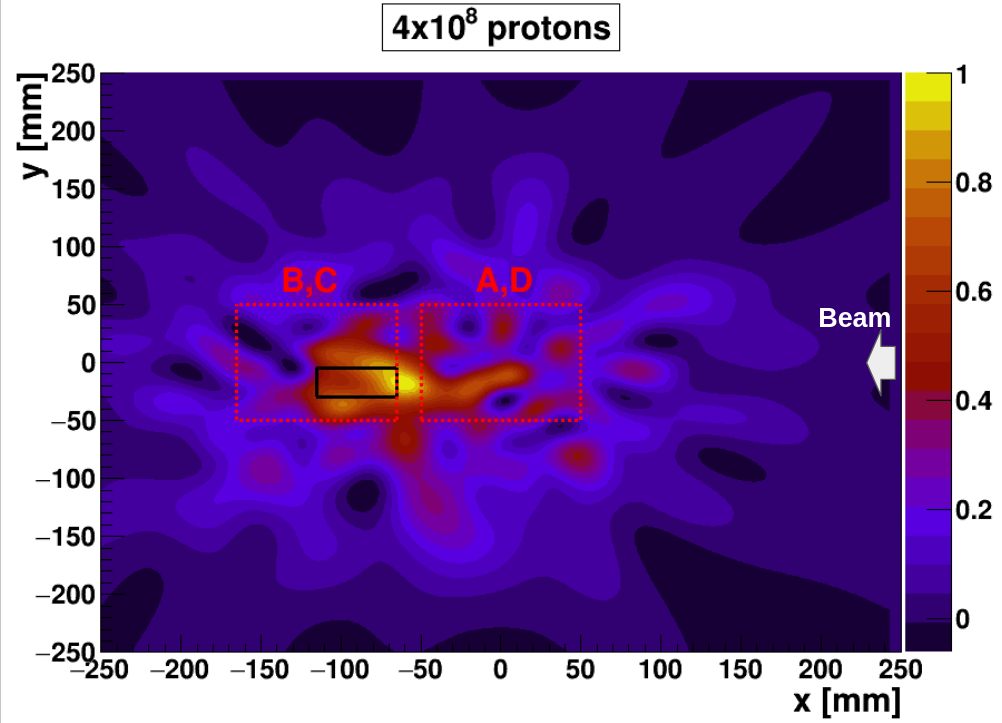} &
\includegraphics[width=0.33\textwidth]{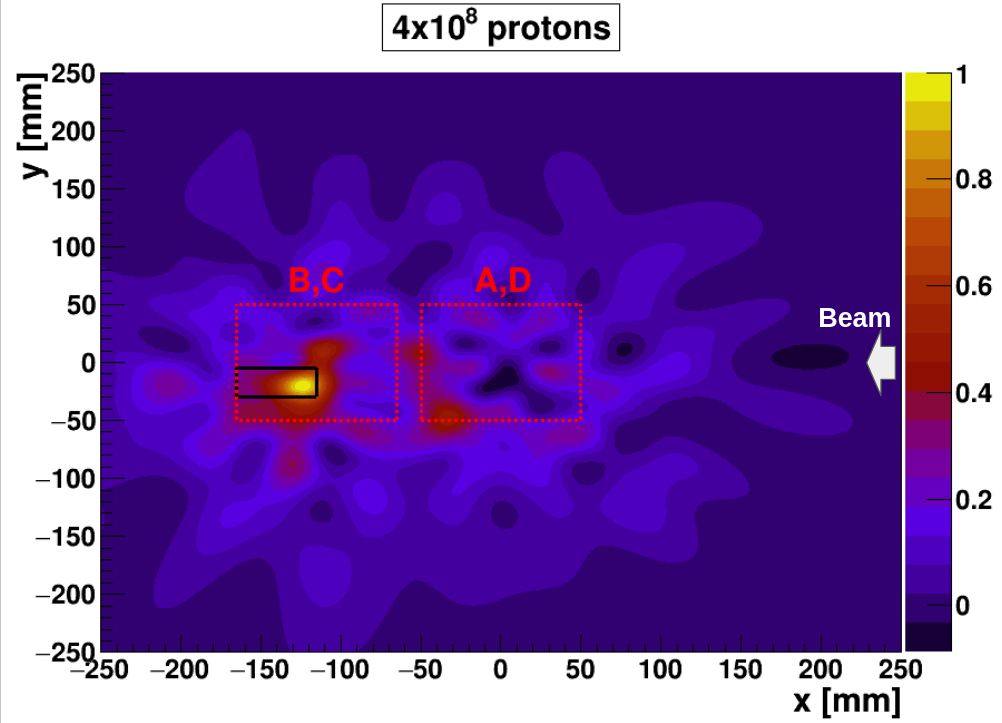} \\
\includegraphics[width=0.33\textwidth]{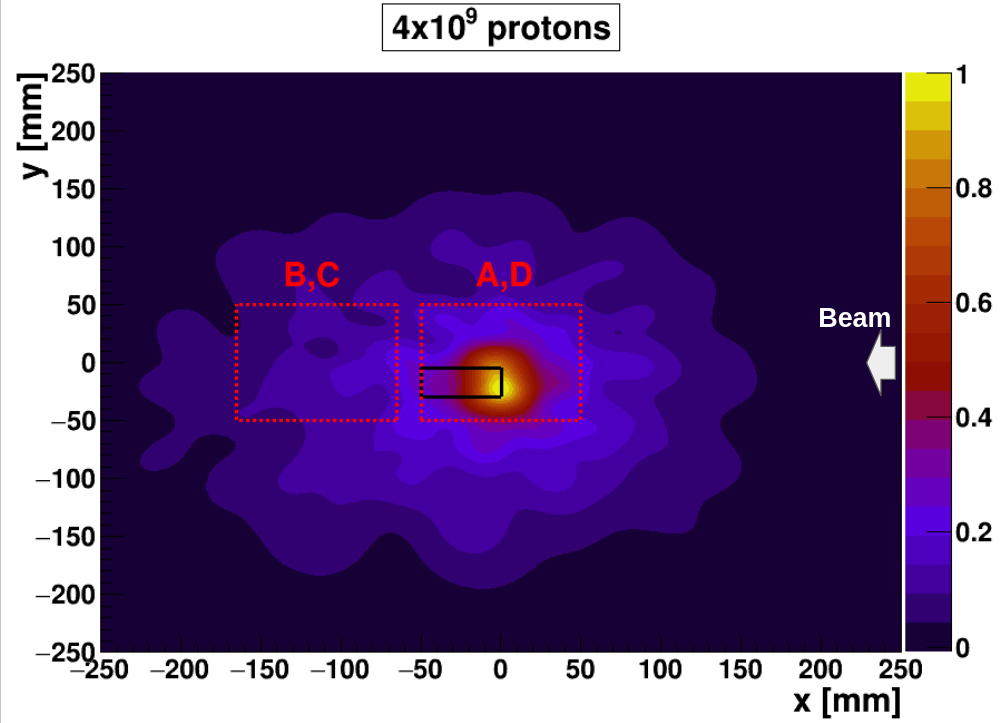} &
\includegraphics[width=0.33\textwidth]{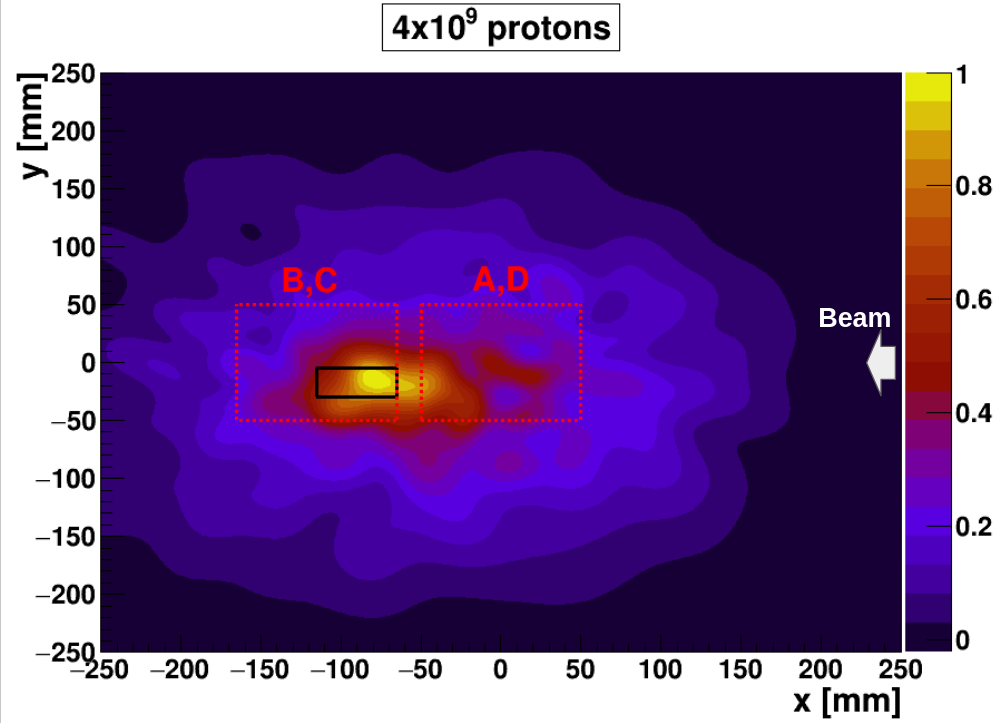} &
\includegraphics[width=0.33\textwidth]{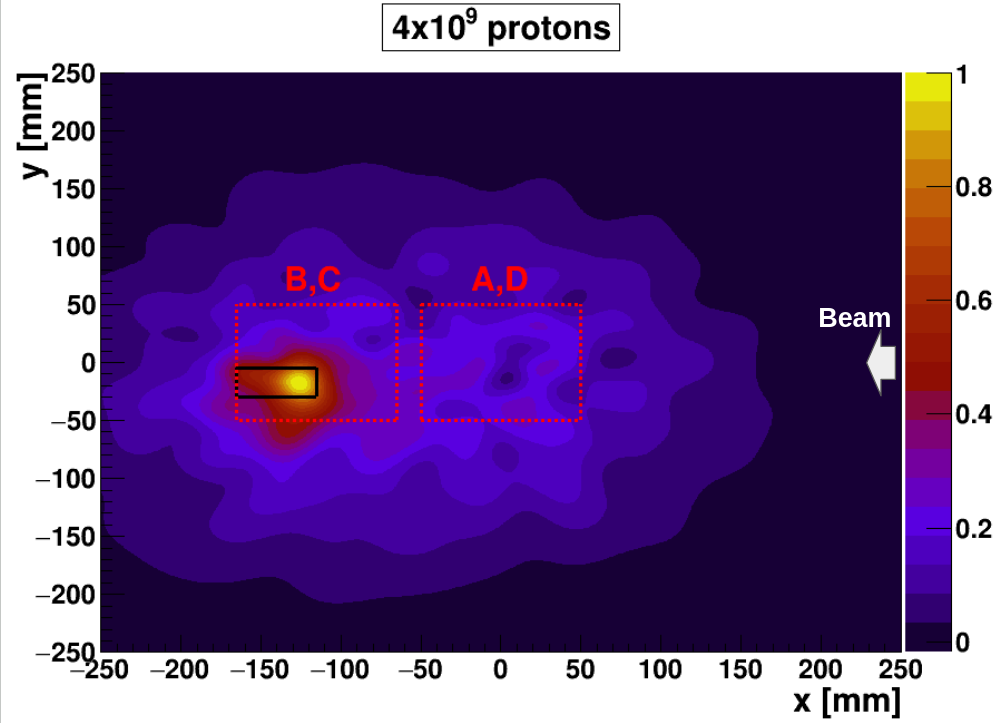} \\ \hline
\end{tabular}
\caption{PGI reconstructed 2D-images for the three different graphite positions and different amounts of cumulative protons. The results presented in the different columns corresponds with the positions presented in Fig.~\ref{fig:fotos} while the rows corresponds to different statistics levels. The contour of the target and Compton-PGI imagers have been overlaid to the results for the sake of interpretation.}
\label{fig:validation}
\end{figure*}

\begin{figure*}[!h]
\centering
\begin{tabular}{c c c}\hline
0 mm & -60 mm & -120 mm \\ \hline \hline
\includegraphics[width=0.33\textwidth]{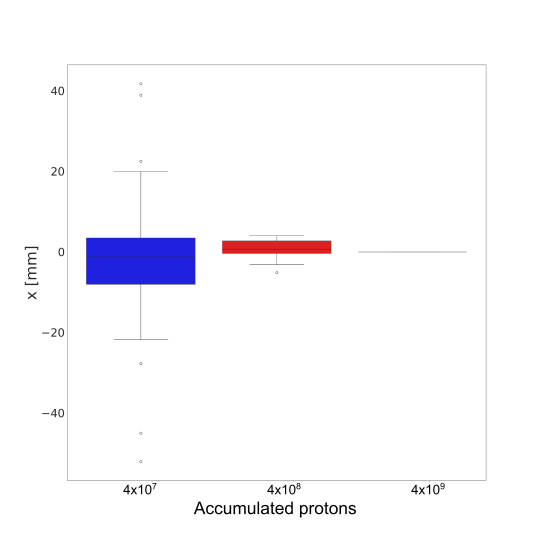} &
\includegraphics[width=0.33\textwidth]{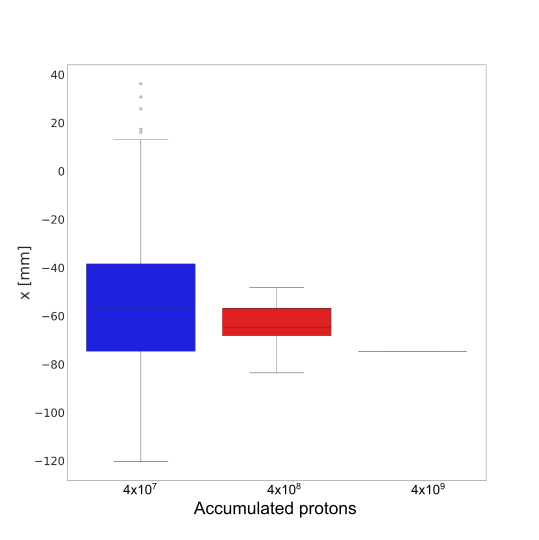} &
\includegraphics[width=0.33\textwidth]{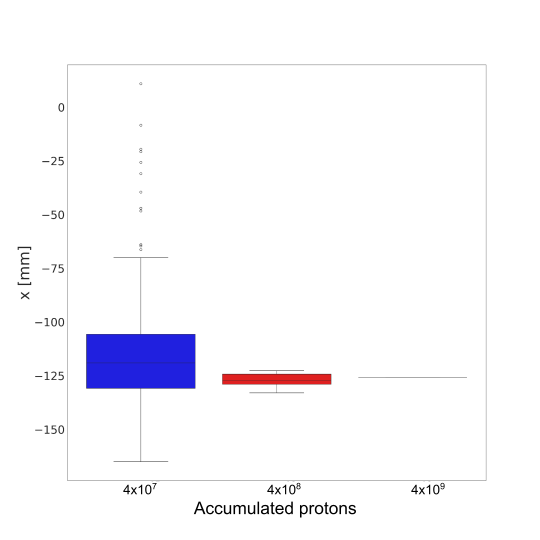} \\
\end{tabular}
\caption{Box plots of the reconstructed graphite $x$-position as a function of different statistics for various graphite positions. From left to right: 0, -60, and -120 mm graphite positions.}
\label{fig:Reconstructed_Pos_Graphite}
\end{figure*}

Important for real-time monitoring in pencil-beam scan is the sensitivity of the distal-edge reconstruction to the number of protons per spot. It is worth noting that $\sim$10$^8$~p is regarded as a clinically relevant quantity, which corresponds to the largest proton delivery per spot in pencil-beam scan for a dose of 1~Gy in a cubic volume of one liter. For Compton-PGI the sensitivity to the Bragg-peak reconstruction in graphite at 55~MeV is shown in Fig.~\ref{fig:validation}, which displays the reconstructed 2D-images for the three graphite target positions and for a different number of accumulated protons on target: 4$\times$10$^{7}$ p, 4$\times$10$^{8}$ p, and 4$\times$10$^{9}$ p. To guide the eye, the contours of the Compton imagers and the graphite cylinder are overlaid on the 2D-reconstructed images. For all reconstructed Compton-PGI diagrams, the maxima of the 2D-reconstructed images are consistent with each other.
From these 2D-diagrams, the Bragg-peak position in the graphite target was calculated as a weighted average using the data points above 95\% of the maximum intensity. The reconstructed $x$-position distributions of the graphite, as a function of the accumulated protons, are presented as box plots in Fig.~\ref{fig:Reconstructed_Pos_Graphite}. As reference, the position for 4$\times$10$^{9}$ protons can be used.
It is important to note that the variance of the distributions decreases by a factor of approximately 3 each time the statistics increases by a factor of 10. Largest deviations are found for the intermediate (-60 mm) position where, as anticipated, the performance of the image reconstruction is affected by the fact that the main locus of $\gamma$-rays lies in the boundary between both pairs of cameras (see Fig.~\ref{fig:fotos}). However, the mean value remains consistent between the reference value (4$\times$10$^9$~p) and those obtained at 4$\times$10$^7$~p and 4$\times$10$^8$~p for all three positions of the graphite target, which lends confidence on the applicability of this technique at clinically relevant intensities ($\sim$10$^8$~p) for proton-beam energies of 55~MeV. It is worth recalling here the relevance of the $\gamma$-ray detection efficiency in this result. The contribution of four (two) Compton cameras to each image reconstruction, as well as their size and geometry (1S+4A)~\cite{Babiano:20} have led to a very high counting statistics already at a rather limited number of protons (10$^{7}$-10$^{8}$~p). However, it is also important to investigate how such large efficiencies could be achieved in a realistic clinical treatment, as well as to envisage an imaging system that is compatible with the clinical environment. Finally, it is worth noting that the results obtained here for 55 MeV protons impinging on the graphite target are comparable to those found in our previous work at the CNA cyclotron at 18 MeV proton beam~\cite{Balibrea22b}, with a distal-edge sensitivity which ranges from sub-millimeter for PET-imaging to a few millimeters for Compton PGI. This situation changes significantly with increasing beam energy, as it will be discused in the following section.

\subsection{Hybrid PET-PGI study with p- and He- beams at 155~MeV and C-ions at 275 MeV}
In the second part of the experiment, a movable 50$\times$50$\times$180 mm$^{3}$ PE target was utilized. Any uncertainty in the positioning of the phantom along the beam axis was eliminated using a high-precision linear stage (see Sec.\ref{sec2}). At variance with our previous study of the hybrid PET-PGI method at low beam energy~\cite{Balibrea22a,Balibrea22b}, here we conducted a broader exploratory survey, including protons, He-ions, and C-ions, as well as beam energies characteristic of clinical treatments (155 MeV for protons and He-ions~\cite{TOMMASINO:2015,Mattei:2017}, and 275 MeV for C-ions~\cite{TOMMASINO:2015,Krimmer:18,Rossi:2022}). With the proton beam, the phantom was placed at three different positions (reference, reference +1 mm, and reference +1.5 mm) along the beam axis, as shown in Fig.~\ref{fig:Exp_setup}. For He- and C-ions, due to the limited beam time available, only two irradiations were feasible (reference and 1 mm shift).
\begin{figure}[!htbp]
    \centering
    \includegraphics[width=0.5\textwidth]{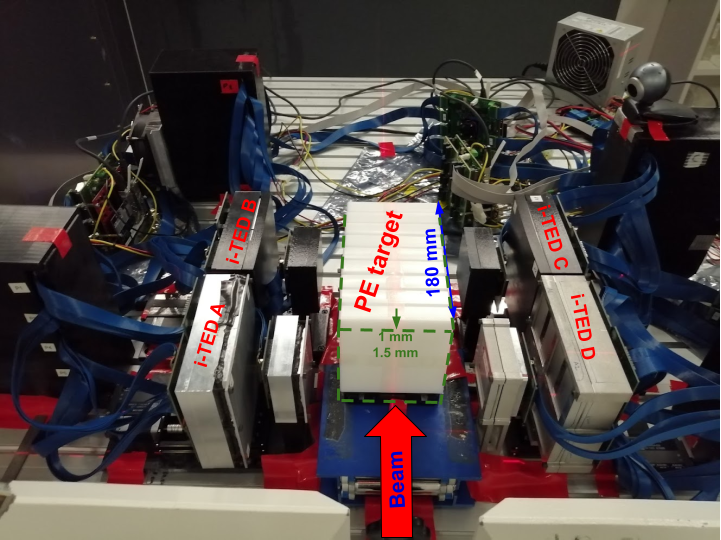}
    \caption{Photograph of the hybrid PET-PGI system, consisting of four Compton imagers in a two-fold front-to-front configuration, taken during the second part of the experiment. The PE target is shown in the center at the nominal position, along with the direction of the 1 mm and 1.5 mm displacements.}
    \label{fig:setup_PE}
\end{figure}

%%%%%%%%%%%%%%%%%%%%%%%%%%%%%%%%%%%%%%%%%%%%%%%%%%%%%%%%%%%%%%%%%%%%%%%%%%%
%
% Discuss expected Bragg position for He and C ions at 155 and 275 MeV ??
% PSTAR not working right now...
%  ME QUEDO AQUÍ, VIERNES 2 AGOSTO 2024
%
%%%%%%%%%%%%%%%%%%%%%%%%%%%%%%%%%%%%%%%%%%%%%%%%%%%%%%%%%%%%%%%%%%%%%%%%%%%

\subsubsection{155~MeV protons}
 
At 155 MeV the Bragg peak was expected to be centered in the FOV of the two downstream Compton imagers (B and C). We conducted three separate pencil-beam scan irradiations of the PE phantom, with a constant transverse beam position at the center of the target and a beam energy of 155 MeV (same layer). For the second and third irradiations the target was shifted by 1000(1) $\mu$m and 1500(1) $\mu$m, respectively, relative to the first (reference) position (see Fig.~\ref{fig:setup_PE}). 
Tab.~\ref{tab:Spills_And_Configuration} in Sec.~\ref{sec2} summarizes the number of spills delivered at each phantom position. The effective number of protons delivered to the target were 1.27×10$^{9}$~p, 5.73×10$^{9}$~p, and 1.27×10$^{10}$~p for the three irradiations or phantom positions, respectively.
%Considering the beam-time structure (see Fig.~\ref{fig:DutyCycle} in {\it Methods}) for PET image reconstruction the total accumulated time in-spill was of 5.71~s, 25.7~s and 57.1~s and the corresponding accumulated time-intervals off-spill were of 28.5~s, 128.9~s and 285.7~s. 
%As it will demonstrate, this is a limiting factor for the sensitivity study with protons.
 
\begin{figure*}[!h]
\begin{tabular}{c c}
\centering
\includegraphics[width=0.5\textwidth]{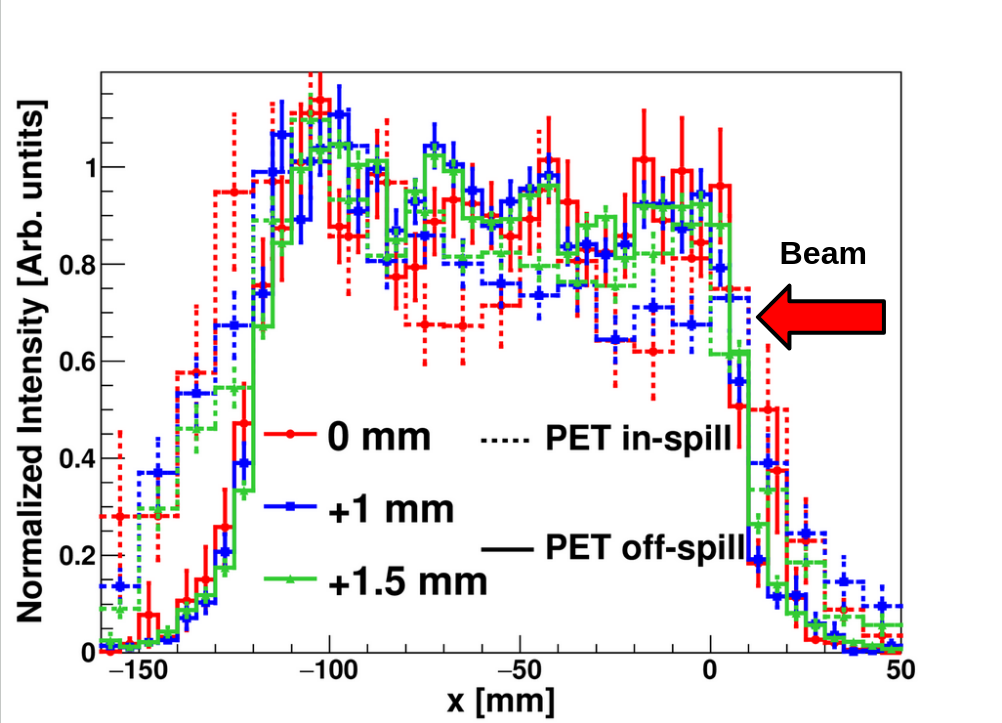} &
\includegraphics[width=0.5\textwidth]{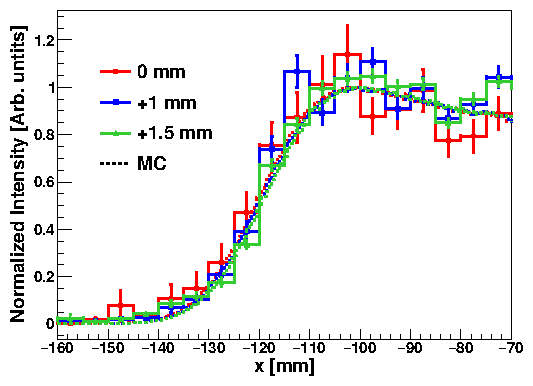} \\
\end{tabular}
\caption{Left: 1D reconstructed PET off-spill (solid) and in-spill (dashed) distributions along the beam axis from the 155 MeV proton beam. Right: Zoom into the fall-off region of the PE off-spill $\beta^{+}$ distribution. The$\beta^{+}$ distributions simulated for the three different positions are shown with a dashed line.} %Left panel: Full Field of View. Right panel: Zoom in on the last few mm.
\label{fig:Protons_PET}
\end{figure*}

PET images were reconstructed using the same methodology as with the graphite target, utilizing 511 keV $\gamma$-rays detected in time coincidence between opposite detectors on each side of the beam axis (see Fig.~\ref{fig:setup_PE}). In-spill and off-spill 1D PET distributions are shown in the left panel of Fig.~\ref{fig:Protons_PET}. As expected for this beam energy, PET reconstructed distributions are rather broad owing to both the activation profile and the large range of the $\beta^+$ particles~\cite{Enghardt:04,Moteabbed:2011,Bauer:13,PARODI:2023}. 

 Fig.~\ref{fig:Protons_PET} shows also a zoom into the fall-off region of the in-beam off-spill $\beta^{+}$distribution. The MC simulated  $\beta^{+}$ activity distribution is also shown for each target displacement. The latter MC calculations have been carried out with the \textsc{Geant4} toolkit~\cite{ALLISON:2016}. A graphical representation of the experimental setup included in the simulation is shown in Fig.~\ref{fig:MC-setup} of Sec.\ref{sec2}. A 155~MeV proton beam with the main HIT synchrotron features~\cite{Bom:2012} was simulated for a total of 10$^{13}$ protons impinging on the PE target. The unstable isotopes produced by the proton interactions were traced when decaying along the proton path, thus reconstructing the $\beta^{+}$ spatial distribution for the different isotopes. The resultant distributions were convoluted with the experimental position resolution in order to be comparable with the experimental reconstructed distribution. Agreement is found between the experimental and MC distributions, consistent with the target position displacement (see below). The MC $\beta^{+}$ spatial distribution is dominated by the contribution of $^{10}$C isotopes, as expected from the synchrotron duty cycle (see Sec.\ref{sec2}) and the large production yield of this isotope with respect to other $\beta^{+}$ unstable nuclei.
 
%In the middle panel the in-beam off-spill distribution is presented along with the Monte Carlo (MC) simulation of the $\beta^{+}$ activity carried out with the \textsc{Geant4} toolkit~\cite{ALLISON:2016}. A graphical representation of the experimental setup included in the simulation is shown in Fig.~\ref{fig:MC-setup} of the {\it Methods} section. A 155~MeV proton beam with the main HIT synchrotron features~\cite{Bom:2012} was simulated for a total of 10$^{13}$ protons impinging on the PE target at the three different positions. The unstable isotopes produced by the proton interactions were traced when decaying along the proton path, thus reconstructing the $\beta^{+}$ spatial distribution for the different isotopes. The resultant distributions were convoluted with the experimental position resolution in order to be comparable with the experimental one. There is relative good agreement between the experimental and MC results. In the same figure there is represented the major contributions to the MC distribution, $^{11}$C, $^{10}$C, and $^{13}$N isotopes, represented in blue, violet, and green, respectively.
As reported in Tab.~\ref{tab:protons_155MeV} the range-shift sensitivity found for the three in-beam off-spill PET distributions is consistent with target-position variations as small as 0.5~mm. The range-shift equivalent was determined from the 50\% PET fall-off position using a Gaussian fit. In contrast, the in-beam in-spill PET distributions are significantly broader compared to the off-spill 1D diagrams. This broadening effect can be attributed to the higher production rates of $e^{+}$ from electromagnetic interactions during proton-beam delivery at high energy, as well as to high-energy $\gamma$-ray interactions (pair production) with the phantom and setup materials. Additionally, the in-spill distributions are forward- (downstream-) peaked, likely due to the increasing contribution of short-lived isotopes. At variance with the situation at 55~MeV, no consistent range-shift trend was found within 1.5 mm target-variations at 155 MeV for the in-spill PET distributions. As a consequence, when increasing the energy by a factor of $\sim$3 only the off-spill 1D PET distributions can be considered reliable.

\begin{table*}[h]
    \centering
    \begin{tabular}{c c c c}\hline \hline
     Configuration & PET off-spill 50\% [mm]  & PGI Max [mm] &  PGI 50\% [mm] \\ \hline \hline
            ref.    & -114.79(20)  & -94(6)  & -161(6) \\
            ref. +1 mm   & -113.96(20) & -90(6) & -158(6)  \\
            ref. +1.5 mm & -113.54(20) & -85(6)  & -168(6) \\ \hline
    \end{tabular}
    \caption{Summary of $x$ range-shift verification results from off-spill PET and PGI using 155~MeV protons. See text for details.}
    \label{tab:protons_155MeV}
\end{table*}

\begin{figure*}[!h]
\begin{tabular}{c c c}\hline
ref. & ref. +1 mm & ref. +1.5 mm \\ \hline \hline
\includegraphics[width=0.33\textwidth]{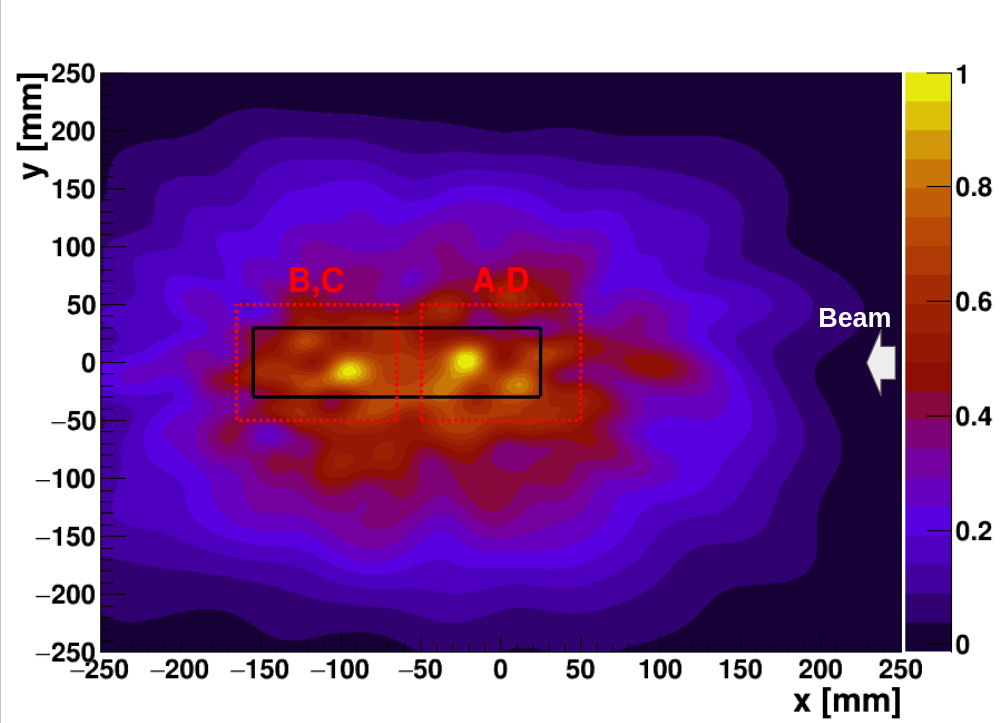} &
\includegraphics[width=0.33\textwidth]{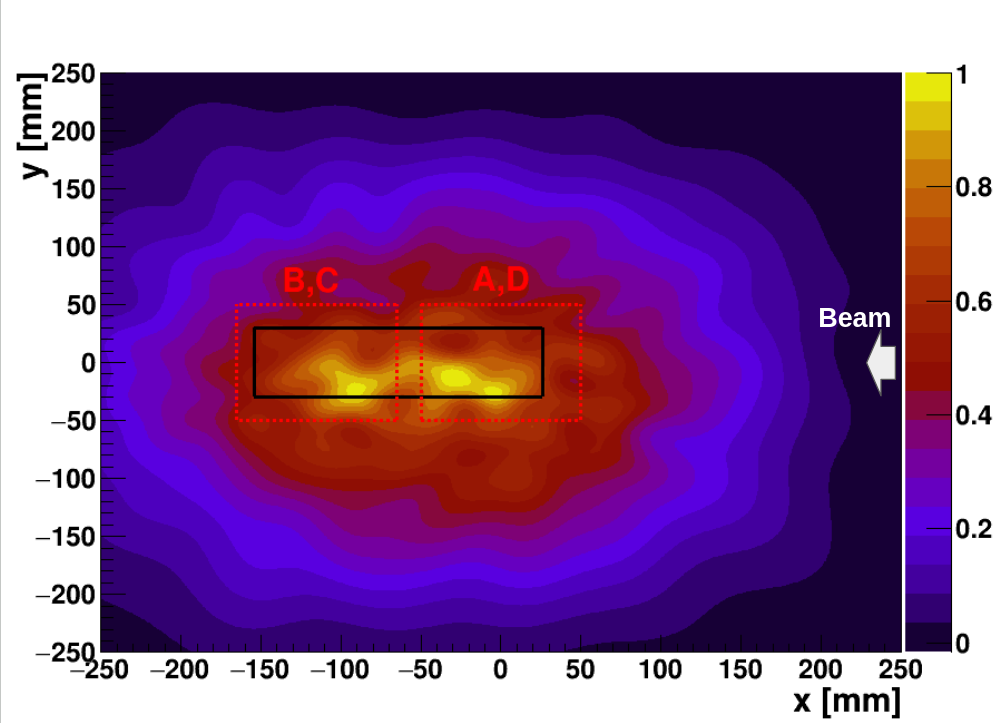} &
\includegraphics[width=0.33\textwidth]{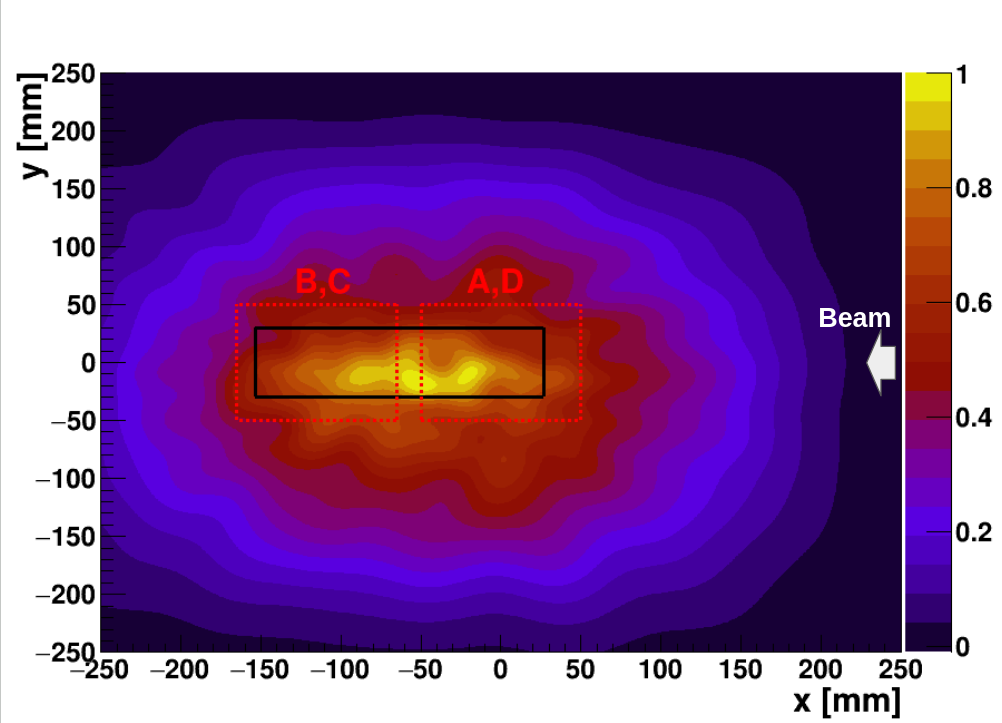} \\
\end{tabular}
\caption{Reconstructed 2D PGI with the 155 MeV proton beam at three different target positions. The phantom is outlined by black lines, and the positions of the various Compton imagers are indicated by red dashed lines. The panels, from left to right, represent the reference polyethylene position and 1 mm and 1.5 mm phantom displacements.}
\label{fig:Protons_PGI}
\end{figure*}

Proton-range variations via PGI become also more challenging at 155 MeV on PE compared to 55 MeV on graphite. The three reconstructed 2D PGI diagrams are depicted in the various panels of Fig.~\ref{fig:Protons_PGI}. Approximately 1-5$\times$10$^4$ events per Compton imager were registered for each phantom position during the irradiations.
 
For all reconstructed 2D diagrams, a relatively high-intensity region is observed around -90 mm, within the central FoV of the B and C imagers. However, the PGI diagrams at this beam energy exhibit irregular shapes, which hinder a reliable range-shift assessment, at least within the explored 1.5 mm phantom position variations, where off-spill PET still performs well (see Tab.\ref{tab:protons_155MeV}). Indeed, as shown in the latter table, for both the maximum and the 50\% fall-off along the beam direction ($x$-axis), there is a lack of agreement between the reconstructed and true positions for PGI within the explored target-position variations of 1.5 mm. Regarding the distribution maxima, for the nominal position, the $x$-axis maximum is located at -94(6) mm, whereas for +1 mm and +1.5 mm, the corresponding maxima positions are at -94(6) mm and -85(6) mm, respectively. This result contradicts the expected trend for the Bragg-peak position. A similar situation is observed when analyzing range-shift variations corresponding to the 50\% fall-off tail.

In order to better interpret the PGI results a series of MC simulations were carried out with the \textsc{Geant4} toolkit~\cite{ALLISON:2016} as presented before. As a consequence of the proton-beam interactions with the PE material a large amount of high-energy prompt $\gamma$-rays~\cite{Moteabbed:2011,Lerendegui:2022} were registered in the Compton imagers. Experimental energy- and position-resolutions~\cite{Babiano:20,Balibrea22b} were included in the simulated detector responses before using them with the Compton-imaging algorithms, which were the same as those utilized experimentally here and in previous works~\cite{Balibrea22b}. The simulated PGI 2D diagrams were analyzed by determining the $x$ (beam-axis) position corresponding to the maximum of each distribution. The spatial sensitivity along the beam axis was then determined from the difference between the maxima for any pair of individual PGI diagrams, $\vert \Delta x_{max}\vert$. From an statistics standpoint the latter quantity corresponds to a permutation test~\cite{Wilcox:2021} and because all MC diagrams were  calculated independently, the cumulative $\vert \Delta x_{max}\vert$ distribution can be interpreted as the probability to determine the Bragg peak position with a precision $\vert \Delta x_{max}\vert$ smaller than a certain target value. 
In short, the cumulative distribution allows one to study the reproducibility and sensitivity of the utilized experimental setup for the implemented image-reconstruction algorithms. The calculation was repeated for different accumulated number of protons per spot, namely 1.0$\times$10$^{9}$, 3.0$\times$10$^{9}$, 6.0$\times$10$^{10}$, and 1.2$\times$10$^{11}$ , which correspond to 10$^{4}$, 3$\times$10$^{4}$, 6$\times$10$^{4}$ and 1.2$\times$10$^{5}$ valid coincidence events per Compton imager.

\begin{figure}[h]
\centering
\includegraphics[width=0.5\textwidth]{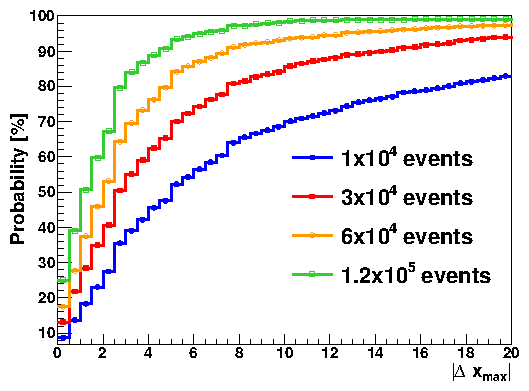} %&
\caption{Cumulative $\vert \Delta x_{max}\vert$ distribution calculated from MC PGI images of the 155 MeV proton beam impinging on the PE target. The color lines represent the different event statistics per individual Compton imager: 10$^{4}$ (blue), 3$\times$10$^{4}$ (red), 6$\times$10$^{4}$ (orange), 1.2$\times$10$^{5}$ (green).}
\label{fig:Deltax}
\end{figure}

The calculated distributions are shown in Fig.~\ref{fig:Deltax}. For the 3$\times$10$^{4}$ events obtained experimentally in each Compton imager with the 155~MeV p-beam on the PE-target after 10$^{9}$-10$^{10}$p in each target position, one can therefore expect a spatial sensitivity $\vert \Delta x_{max}\vert$ not better than 14~mm at 90\% confidence level. Almost a factor of two increase in Compton efficiency could be obtained by means of a four-fold Compton set-up, such as the one shown in Fig.~1 of Ref.~\cite{Lerendegui:2022}. For the latter geometry the expected fall-off retrieval precision could be increased to about 8~mm at 90\% confidence level for 10$^9$ protons on target. This is still one order of magnitude larger than the $\sim$10$^8$~p  relevant for quasi-real time monitoring in clinical treatments. %Alternatively, ten times more protons on target or spot ($\sim$10$^{10}$ protons) would be necessary with the present setup to enhance the number of Compton events for PGI to $\sim$10$^{5}$ and thus achieve a fall-off retrieval position precision of about 4~mm at 90\% confidence level. 
%In the same sense, increasing the statistics by a factor two or four, the reproducibility is decreased to 8~mm and 4~mm, respectively at the same confidence level. The same argument is attainable to the 50\% fall-off of the reconstructed PGI distribution, usually attainable to the Bragg peak position reconstruction~\cite{Lerendegui:2022}.
In summary, while the combined PET-PGI fall-off retrieval precision at 55~MeV was found rather consistent and satisfactory with both techniques, at high proton beam energy (155~MeV) the results obtained here indicate that only in-beam off-spill PET seems to provide enough sensitivity towards sub-mm range verification in real time. 

\subsubsection{155~MeV He-ions}
The PE-target was shifted 20~mm upstream in order to keep the expected Bragg-peak position for the 155~MeV He-beam in the central FOV region of the B- and C-imagers. Due to the limited beam-time availability only two exploratory phantom irradiations were carried out, which corresponded to a reference phantom position and a 1~mm shift downstream. 
On average, 1085 spills and $\sim$5$\times$10$^{9}$ He-ions were delivered in each irradiation or target position.
%and 57/285 effective seconds time periods for in- and off-spill PET reconstruction.

\begin{figure*}
    \centering
    \begin{tabular}{c c}
    \includegraphics[width=0.5\textwidth]{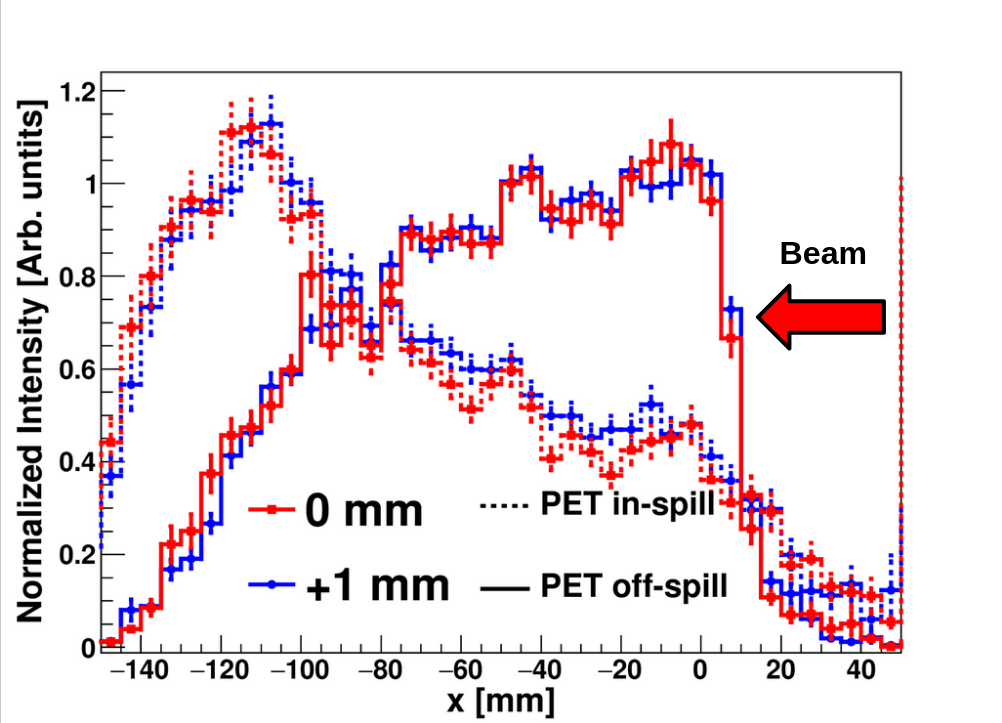} &
    \includegraphics[width=0.5\textwidth]{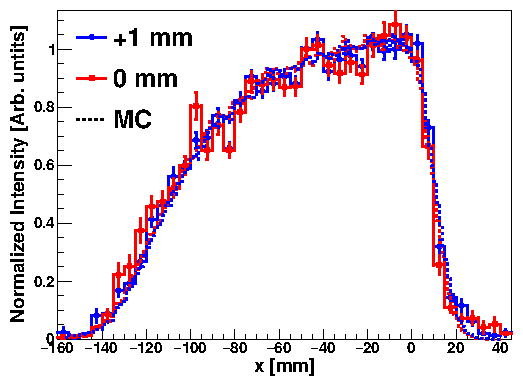}\\
    \end{tabular}
    \caption{Left: PET 1D distributions reconstructed in-spill (dashed lines) and off-spill (solid lines) for the 155~MeV He-ion beam. Right: Comparison between PET 1D off-spill distributions measured for a target-shift of 1 mm. Corresponding MC activity distributions are also shown with dashed lines.}
    \label{fig:PET_alpha}
\end{figure*}

Fig.~\ref{fig:PET_alpha} shows the reconstructed in-beam in- and off-spill PET 1D distributions. Since only two data points (positions) were available, only a quantitative interpretation of the results was carried out, with no further effort to perform a systematic assessment of the ion-range sensitivity from this data. Interestingly, the in-spill PET distributions are peaked toward the end of the phantom, with a maximum in both irradiations approximately at the position where the Bragg peak is expected. On the other hand, the off-spill PET distribution shows the opposite trend, indicating that short-lived isotopes may be predominantly produced in the vicinity of the Bragg peak, while long-lived nuclei exhibit a relatively constant production yield across the target volume, which decreases as the beam energy decreases. The $\beta^{+}$ profile determined from the in-beam off-spill distribution is consistent with the fact that stable beams of $Z<$5 can produce target fragments with significant positron emission yields all along the primary beam penetration~\cite{PARODI:2023}. The off-spill PET distribution is compared in Fig.~\ref{fig:PET_alpha} against MC simulations of the $\beta^{+}$ activity distribution calculated with \textsc{Geant4}~\cite{ALLISON:2016}. The MC simulation was performed as explained in the previous section, replacing here the protons by 155 MeV He-ions. The experimental distribution can be reconstructed only using the calculated $^{9}$C (t$_{1/2}$ = 126.5 ms) spatial distribution. This can be ascribed to a larger sensitivity to short-lived isotopes because of the chosen synchrotron duty-cycle and the  $^{9}$C large production yield compared with the rest of short-lived isotopes, as indicated by the MC simulation. It is worth mentioning that the trend of the experimental distributions is in agreement with the MC simulation, consistent with a displacement of the distribution by 1~mm.

\begin{figure*}[h]%
%\centering
\begin{tabular}{c c c}\hline
ref. & ref. +1 mm & MC \\ \hline \hline
\includegraphics[width=0.33\textwidth]{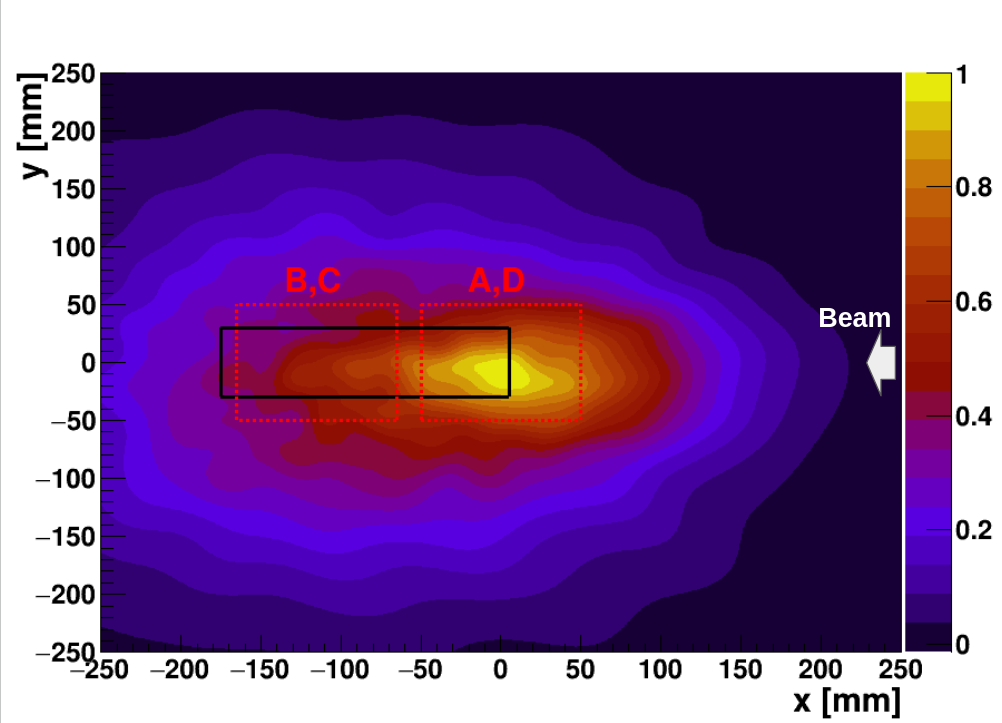} &
\includegraphics[width=0.33\textwidth]{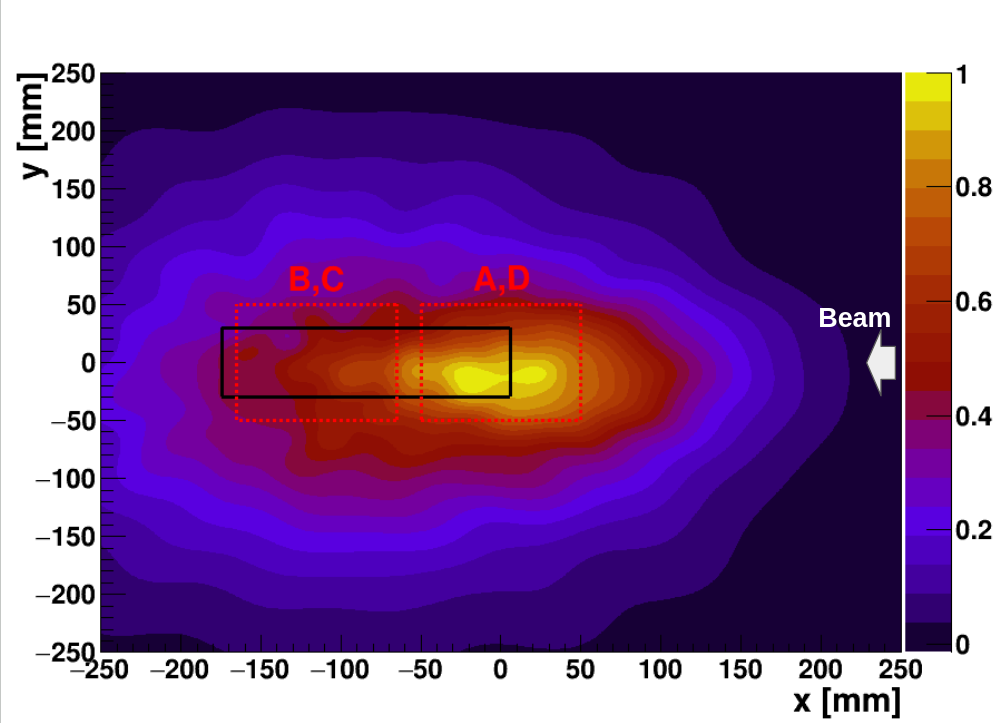} &%\\
%\end{tabular}
%\centering
\includegraphics[width=0.33\textwidth]{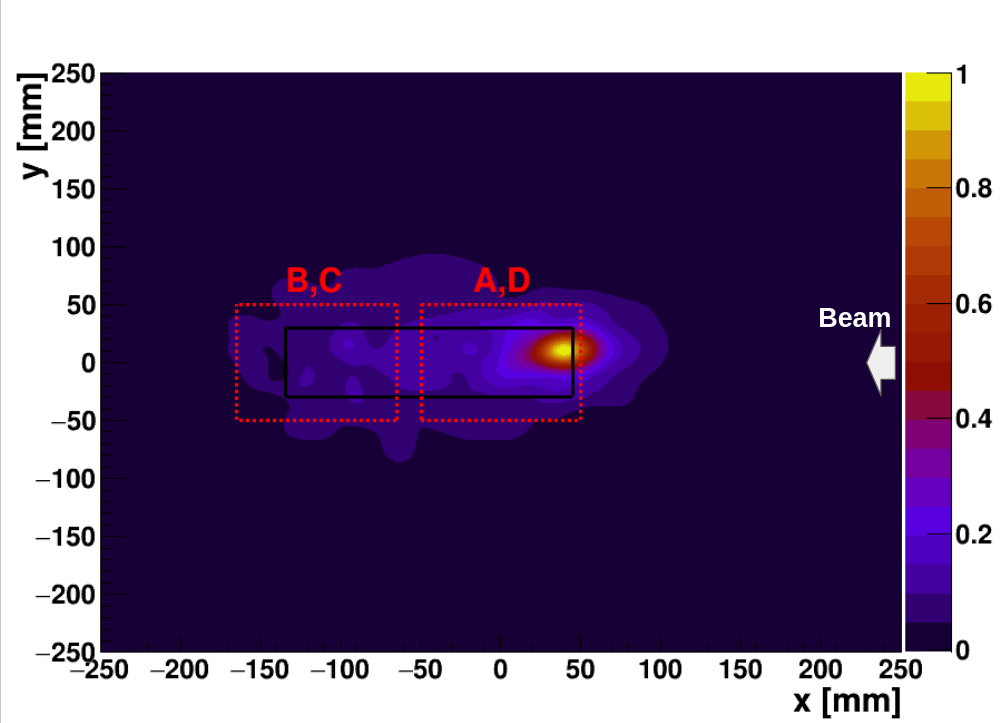} \\
\end{tabular}
\caption{2D PGI reconstructed diagrams for the 155~MeV He-ion beam. From left to the right, nominal phantom position, 1~mm displacement, and MC reconstructed PGI diagram for the experiment realization.}
\label{fig:PGI_alpha}
\end{figure*}

Reconstructed Compton-PGI 2D diagrams for the He-ion beam are shown in Fig.~\ref{fig:PGI_alpha}. The images show a strong radiation locus centered in the FOV of detectors A and D, which obviously does not coincide with the Bragg-peak position that is expected in the central FOV of detector imagers B and C. 
In order to interpret the PGI result, a series of MC simulations were carried out in a similar fashion as described before. In this case, a beam of He-ions at 155 MeV was simulated, impinging in the center of the PE-phantom. The results from the MC calculation are also displayed in the right panel of Fig.~\ref{fig:PGI_alpha} and show a sort of strong $\gamma$-flash at the entrance position of the beam in the target. The simulated $\gamma$-ray pattern shows a reasonable agreement with the radiation pattern determined experimentally, which is also shown in the same figure. Inspection of the MC events indicates that, owing to the higher Coulomb barrier of the He-ions, and the relatively large nuclear cross sections at 155 MeV several nuclear-reaction channels are readily open at the entrance path of the beam in the target. This includes fragmentation reactions, proton- and neutron knock-out reactions, etc. All these nuclear interactions emit prompt $\gamma$-rays, which hinder the observation of the $\gamma$-rays emitted later, shortly before the Bragg-peak. 

In summary, in a similar situation as for 155 MeV proton-beams, hybrid PET-PGI with He-ions at 155 MeV seems to be rather limited to only in-beam off-spill PET ion-range assessment, at least at the level of clinical intensities of 10$^8$~p per spot. It remains to be investigated in future studies the possibility to perform PET-PGI at lower He-ion beam energies and to determine the energy regime, where both imaging approaches can be simultaneously exploited. Further possible upgrades in the detection system, aimed at enhancing the performance of PGI for treatments with He-ions need to be also further investigated.

\subsubsection{275~MeV C-ions}
Similarly to previous configurations, the average number of spills per irradiation was of 1002, which correspond to an average of $\sim$5$\times$10$^{9}$ C-ions delivered to the target. A phantom separation of only 1~mm between both irradiations was investigated.  
\begin{figure*}[!htbp]
    \centering
    \begin{tabular}{c c}
    \includegraphics[width=0.5\textwidth]{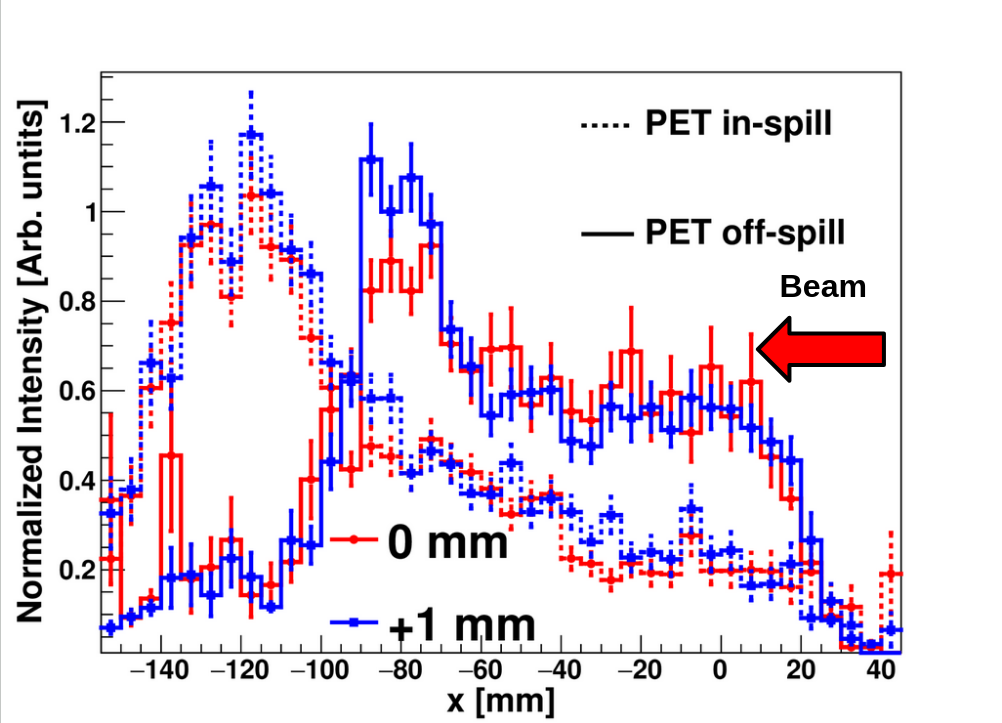} &
    \includegraphics[width=0.5\textwidth]{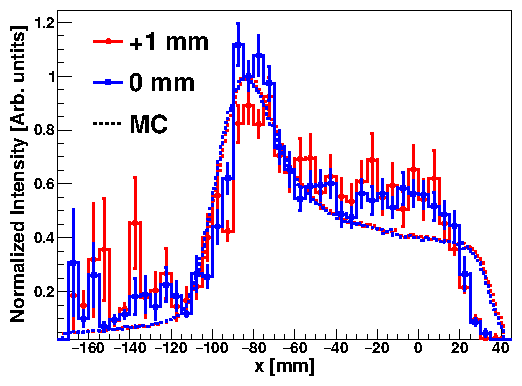} \\
    \end{tabular}
    \caption{Left: 1D in-spill and off-spill PET images reconstructed for 275~MeV C-particles beam on a PE target. Nominal and 1 mm shifted phantom configurations are presented by red and blue lines, respectively. Right: In-beam off-spill experimental distributions for a target-shift of 1mm and corresponding MC calculations of the $\beta^{+}$ activity distribution along the beam axis.}
    \label{fig:PET_C}
\end{figure*}
Fig.~\ref{fig:PET_C} shows the in-beam measured PET distributions in-spill and off-spill. The in-spill PET distribution shows a similar situation to the one found before with the He beam. The Lorentz boost from highly energetic C-ions induces a large contribution of $\beta^{+}$ emitters downstream, giving rise to a prominent peak at around -120~mm. Within the statistical accuracy of the measurement there is no possibility to disentangle mm-size range-shift variations from the in-spill PET distributions. The off-spill PET projections show a small contribution from relatively long-lived $\beta^{+}$-emitters beyond -100~mm, which can be ascribed to projectile-fragments reaching beyond the C-ion Bragg peak~\cite{PARODI:2023,Fischetti:2020}. The off-spill bulk of $\beta^+$ strength is concentrated at -80~mm, which coincides with the expected Bragg-peak position for the primary C-beam. From -60~mm up to 20~mm a relatively flat $\beta^{+}$ contribution is found, which can be interpreted as the activation of relatively long-lived $\beta^{+}$ emitters  along the C-ion trajectory through the phantom volume. The trend of the experimental data is well reproduced by the MC simulations of $^{9}$C activity, as shown in the right panel of Fig.\ref{fig:PET_C}. Deviations between measured and calculated profiles can be attributed to the relatively poorly known $\beta^{+}$ production cross section for this reaction in this energy regime. The experimental off-spill $\beta^{+}$ activity profile calculated in this figure is consistent with previous studies conducted by other research groups \cite{PARODI:2023,Fischetti:2020}. It is worth noting that even with a factor of $\sim$100 higher number of C-ions than the clinically relevant value ($\sim$10$^7$ C) and with high-efficiency detectors as those utilized here, the counting statistics still hinder an assessment of the 1~mm range shift. 

%In this case, Bragg peak is clearly observed in off-spill PET distributions because of large amount of C isotopes deposited at Bragg peak positions, as expected. In previous cases, neither the proton or alphas can be stripped, thus the Bragg peak position can not be determined directly from the off-spill PET distributions. 

\begin{figure*}[h]%
\centering
\begin{tabular}{c c c}\hline
ref. & ref. +1 mm & MC \\ \hline \hline
\includegraphics[width=0.33\textwidth]{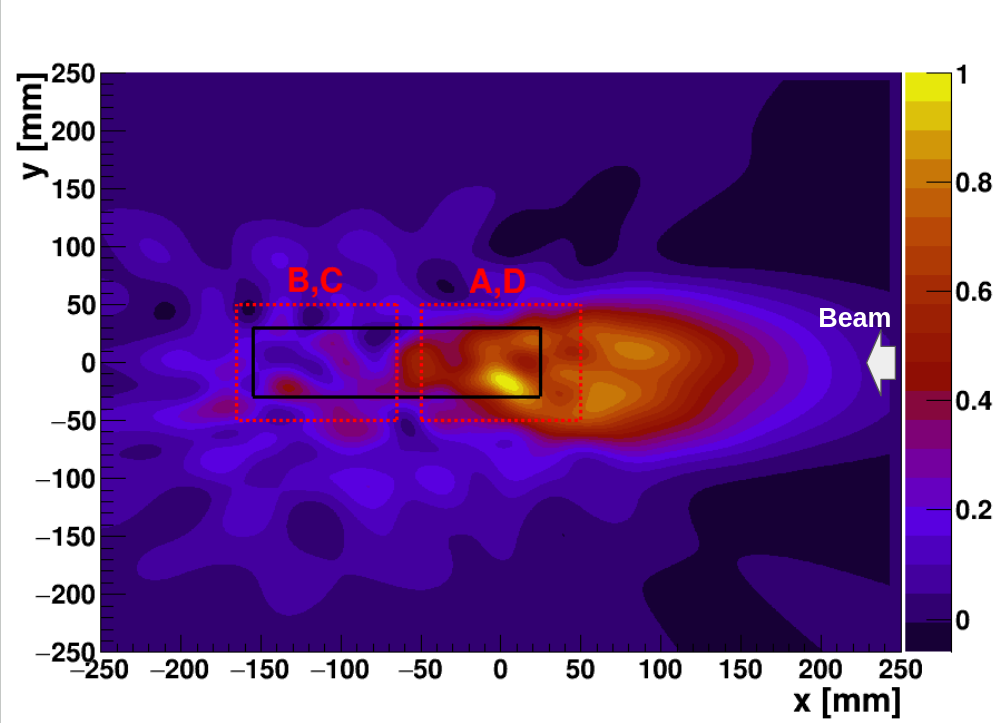} &
\includegraphics[width=0.33\textwidth]{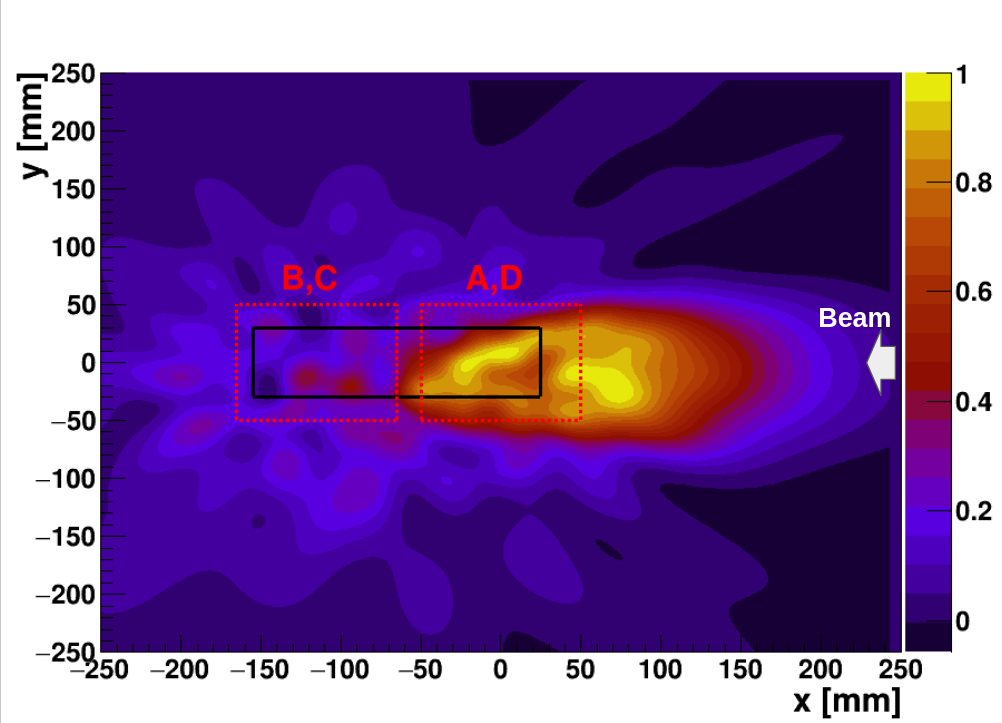} &%\\
%\end{tabular}
%\centering
\includegraphics[width=0.33\textwidth]{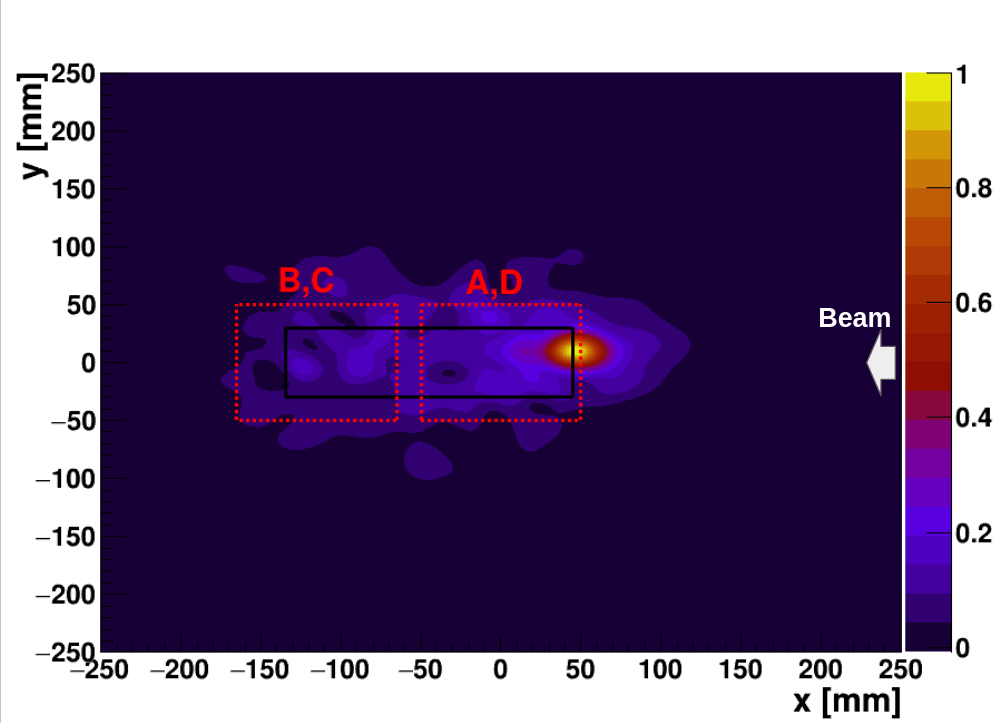} \\
\end{tabular}
\caption{2D PGI reconstructed diagrams for 275~MeV C-ion impying on the PE target. From left to the right, nominal PE phantom position, 1~mm displacement, and MC reconstructed PGI diagram using the same experimental conditions.}
\label{fig:PGI_C}
\end{figure*}

The experimental and MC PGI reconstructed diagrams for the C-ion irradiations are shown in Fig.~\ref{fig:PGI_C}. In a similar situation as with the He-beam, the bulk of high-energy prompt $\gamma$-ray strength is concentrated at the beginning of the PE phantom, thus rendering the prompt $\gamma$-rays in the vicinity of the Bragg peak essentially unnoticeable. Thus, the effect of this flash of $\gamma$-rays at these projectile energies seems to be a challenging factor for real-time ion-range verification via PGI utilizing He- and C-ions. As discussed before, it would be of great interest to explore other beam-energy regimes, as well as possible setup upgrades, which may help to retrieve information about the Bragg-peak position via imaging of prompt $\gamma$-rays. 
%After these measurements, in-beam off-spill planar PET imaging seems the most promising technique for online ion-range verification using $\alpha$- and C-ions at relatively high beam energies. 

%In the same sense as for $\alpha$-particles, the nuclear reactions are taking at the entrance where the projectile energy because of the larger number of nuclear reaction channels opened while the energy deposition is taking mainly at the Bragg peak. It is important to remember that prompt $\gamma$-ray emission takes place at the position where the nuclear reaction takes place and it could be different from the energy deposition. The obtained results are in the direction of use heavy shielding for prompt  $\gamma$-ray based range verification at these very high energies as it was stated in previous works~\cite{Min:2006,Testa:2006,Pinto:2015}.

%As in previous cases, MC PGI diagrams using the same conditions were reconstructed in order to validate the current hypothesis. In this case, we simulated C$^{12}$ ions at 275~MeV with the same experimental distribution. The results, shown in the right panel of Fig.~\ref{fig:PGI_C}, are in agreement with the experimental observation and support the aforementioned hypothesis: The nuclear reactions emitting prompt $\gamma$-rays are taking place vastly at the entrance of the phantom compare to those occurring near to the Bragg peak position.

\section*{Summary and conclusion}
%Prompt-gamma Compton imaging and positron-emission tomography show complementary features and a system capable of combining them both simultaneously could be of interest for enhanced accuracy ion-range verification in hadron therapy~\cite{Parodi:16}. 
In this work we have experimentally explored the concept of hybrid Compton-PET imaging~\cite{Parodi:16} for the first time at clinical-beam conditions with the aim of exploring its feasibility for ion range verification. 
The proof-of-concept measurements were carried out at HIT-Heidelberg. For convenience, the beam-time structure of the HIT synchrotron was tuned for a pulsed-beam (45~ms on and 255~ms off), which is especially well suited for investigating ion-range verification via three modalities at the same time: PGI, in-beam in-spill PET and in-beam off-spill PET. The experimental set-up consisted of four Compton imagers that were placed in a twofold front-to-front configuration to cover the entire irradiated phantom simultaneously by the Compton- and PET- field of views. This configuration enabled a rather complete overview of the beam-interactions in the target and it was of particular interest for properly interpreting the interactions of particles heavier than protons.  

In a first run of measurements a cylindrical graphite target was placed at three different positions along the beam axis and irradiated with 55~MeV protons. Although 55 MeV is relatively low in comparison to most clinical treatments, it is commonly applied in the treatment of uveal melanoma, conjunctival melanoma, and other malignant and benign ocular pathologies. 
The results obtained for the full statistics ($\sim$4$\times$10$^9$ p) show an excellent agreement between both PGI and PET image-reconstruction methodologies, within the corresponding systematic uncertainties. At this beam energy position-reconstruction via PET becomes better than 1~mm (both in-spill and off-spill), whereas deviations of 3-4~mm are found with Compton-based PGI. 

A sensitivity study was made to asses the feasibility of real-time range verification via PGI. A sensible quantity is 10$^8$ accumulated protons, which correspond to highest intensity spot in a conventional clinical treatment. Thus, PGI was applied for different values of accumulated protons spanning from 4$\times$10$^7$ up to 4$\times$10$^9$ p. The resulting reconstructed position distributions  indicate that while the mean values of the distributions remain compatible, the variance decreases by a factor of approximately 3 for every tenfold increase in the number of protons. Satisfactory results were found for 4$\times$10$^8$~p. This value could still be improved (reduced) in the future after some upgrades in the geometry of the experimental setup.

In a second series of runs, higher beam energies and heavier ions were utilized. The graphite target was replaced by a series of large PE blocks. Protons, He- and C-ions were delivered to the target.

\begin{figure}[!h]
\centering
\includegraphics[width=0.5\textwidth]{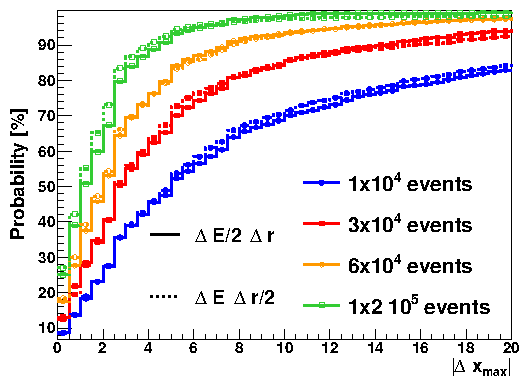}% \\
\caption{$\vert \Delta x_{max}\vert$ cumulative distribution for different target statistics increasing the deposited energy resolution by a factor 2 (solid) and by reducing the position reconstruction uncertainty by a factor 2 (dashed).}
\label{fig:DeltaxDiss}
\end{figure}

In the first series of measurements, 155~MeV protons were used with 3 different PE target positions separated by 1 and 1.5~mm along the beam axis from the nominal phantom position. Reconstructed 2D PGI diagrams show a main locus of $\gamma$-rays at a beam-axis position of -100~mm, which matches well with the expected position for the Bragg peak. However, position-reconstruction accuracy was significantly worse than at lower (55 MeV) energy. This result could be understood on the basis of MC simulations and permutation tests, which indeed revealed an expected accuracy not better than 14~mm at 90\% confidence level for PGI. In order to explore possible future optimizations, the MC sensitivity calculation was repeated for energy- and position-resolution values artificially improved by a factor 2. As shown in Fig.~\ref{fig:DeltaxDiss}, the Bragg peak position reconstruction is vastly dominated by statistics rather than systematics (intrinsic resolutions). Therefore, we can conclude that in order to improve the sensitivity of PGI further, the geometrical configuration must be optimized to increase the efficiency in the vicinity of the Bragg peak, where the physics signals are more pronounced. A cross-shape configuration with four Compton arms, similar to the one proposed in a previous work~\cite{Lerendegui:2022}, could provide an enhancement in efficiency by a factor of two. This intriguing result raises the important question of whether a PGI system that is sufficiently efficient, yet necessarily bulky, can be compatible with the clinical environment.

In- and off-spill 1D PET distributions for 155 MeV protons on the PE-target show a relatively broad distribution. The reconstructed in-beam off-spill PET distribution yields a sub-mm accuracy, which is comparable with results reported by other groups utilizing dedicated large planar PET systems~\cite{PARODI:2023,Ozoemelam:2020a}.

At 155~MeV we also investigated the performance of the hybrid PET-PGI system with He-ions. Interestingly, in this case the largest high-energy gamma-ray emission takes place at the beginning of the ion-path through the target, thereby hindering a reliable assessment of the Bragg-peak position. This result was fully consistent with the reconstructed PGI diagrams from MC simulations. 1D in-spill PET distributions offer a complementary view of the same process in which projectile fragments are drifted towards the end of the phantom, even beyond the Bragg peak estimated position. The in-beam off-spill PET distributions were found the most suitable imaging technique for ion-range verification with $\alpha$-beams of 155~MeV. This result is in agreement with other works published recently~\cite{Maccabee:1969,Ozoemelam:2020a,PARODI:2023}. 
%It is worth to mention that both, in- and off-spill PET distributions observe the 1~mm shift between measurements.

Finally, a short test was conducted utilizing C-ions at 275~MeV with a large PE target at two different positions separated by 1~mm. In a similar situation as with He beams, there is a strong prompt $\gamma$-ray emission at the entrance of the PE phantom, which hinder the applicability of PGI for inspecting the Bragg peak. On the other hand, in-beam off-spill PET distributions deliver best performance in terms of sensitivity to the phantom shift.

In summary, in this first exploratory study at clinical-beam conditions, hybrid PET-PGI has been successfully implemented and it has been found to be a promising technique for accurate real-time monitoring ($\sim$10$^8$ p) at relatively low clinical energy (55~MeV). However, the performance of the PGI part seems to become increasingly challenged with higher beam energies (155~MeV or more) and with heavier beam particles (He and C ions). One of the most interesting aspects to be investigated in future works is the regime of energies, namely between 55~MeV and 155~MeV, where one could still apply the hybrid imaging concept, especially for extensively used proton beams. Another topic to be researched is the possibility to implement new features in the detection system, which mitigate the strong effect of the $\gamma$-flash produced at the entrance of the target at high beam energy and with heavy particles. Implementing a solution for the latter effect could help to extend the hybrid concept to a broader range of clinical situations and treatments. Finally, we have seen the relevance of sufficient counting statistics for accurate ion-range assessment in real time. In spite of utilizing Compton imagers with unparalleled detection efficiency in this field, even for an optimized system where four imagers are pointing to the Bragg-peak position, it remains difficult for PGI to reach enough statistics for real-time range accuracies of $\lesssim$1~mm with clinical intensities (10$^8$ protons per spot). Therefore, further upgrades and highly optimized designs need to be envisioned when aiming at online monitoring, while still preserving compatibility with the clinical environment.

\section*{Acknowledgements}
This work derives from fundamental research carried out with funding from the European Research Council (ERC) under the European Union's Horizon 2020 research and innovation programme (ERC Consolidator Grant project HYMNS, with grant agreement No.~681740) and part of this work has been funded under ERC Proof-of-Concept Grant GNVISION (Grant Agreement 101113330). The authors acknowledge support from the Spanish Ministerio de Ciencia e Innovaci\'on under grants PDC2021-121536-C21, PID2019-104714GB-C21 and RTI2018-098117-B-C21. We acknowledge support from the Severo Ochoa Grant CEX2023-001292-S funded by MCIU/AEI. This work was partially supported by Generalitat Valenciana PROMETEO/2019/007. T. Rodriguez-Gonzalez acknowledges the Spanish FPI predoctoral grant. J. Balibrea-Correa is supported by grant ICJ220-045122-I funded by MCIN/AEI/ 10.13039/ 501100011033. J. Lerendegui-Marco is supported by grant FJC2020-044688-I and CIAPOS/2022/020 funded by MCIN/AEI/ 10.13039/ 501100011033 and Generalitat Valenciana, respectively. M.C. Jiménez-Ramos acknowledges the support to this work through a VI PPIT-US contract. We would like to thank the crew at the Electronics Laboratory of IFIC, in particular Manuel Lopez Redondo and Jorge N\'acher Ar\'andiga for their excellent and efficient work prototyping and manufacturing customized PCBs for this work. 
%Acknowledgements should be brief, and should not include thanks to anonymous referees and editors, or effusive comments. Grant or contribution numbers may be acknowledged.

\section*{Author contributions statement}
JB-C contributed to investigation, methodology, formal analysis, data curation, visualization, and writing-original draft. JL-M
contributed to investigation, methodology, formal analysis, and data curation. IL contributed to hardware, software, and visualization. SM contributed software, and visualization. CG contributed to conceptualization, methodology, supervision, investigation, and formal analysis. TR-G contributed to investigation, methodology, formal analysis, and data
curation. MCJ-R contributed to conceptualization, methodology, supervision, investigation, and formal analysis. JM-Q contributed to conceptualization, methodology, supervision, investigation, and formal analysis. JB contributed to conceptualization, supervision, investigation, and formal analysis. SB contributed to conceptualization, supervision, investigation, and formal analysis. CD-P contributed to conceptualization,
methodology, supervision, writing-review editing, funding acquisition, investigation, and formal analysis.

\section*{Data availability statement}
Data will be available on reasonable request.
\section*{Additional information}

\textbf{Competing interests} The authors declare no competing interests. 
%
% BibTeX users please use
 \bibliographystyle{unsrt}
 \bibliography{sample}
%\begin{thebibliography}{}
%Author, \textit{Book title} (Publisher, place year) page numbers
% etc
%\end{thebibliography}

\end{document}